\documentclass[11pt,a4paper]{article}
\pdfoutput=1

\IfFileExists{srcltx.sty}{\usepackage[active]{srcltx}}

\usepackage[final]{draftwatermark} 
\SetWatermarkScale{7.0}
\SetWatermarkLightness{0.90}

\usepackage{epsfig}
\usepackage{amsfonts}
\usepackage{bbm}
\usepackage{lineno}
\usepackage{units,slashed}
\usepackage{amsmath,amssymb} %
\usepackage{xspace,bm,paralist,graphicx,color}

\usepackage{hyperref}    % Hyperlinks in references

\usepackage{cite}

\def\be{\begin{equation}}
\def\ee{\end{equation}}
\def\ba{\begin{eqnarray}}
\def\ea{\end{eqnarray}}
\def\lsi{\raise0.3ex\hbox{$<$\kern-0.75em\raise-1.1ex\hbox{$\sim$}}}
\def\gsi{\raise0.3ex\hbox{$>$\kern-0.75em\raise-1.1ex\hbox{$\sim$}}}

\newcommand{\gev}{\ensuremath{\;\mathrm{GeV}}} %
\def\d{\partial}

\begin{document}

%\pagestyle{empty}

%\begin{titlepage}
%\begin{flushright}
%BI-TP 2007/xx\\
%\date{today}
%\end{flushright}
%\begin{centering}
%\vfill

%\title{
{
\begin{flushright}
\textnormal{\normalsize ~}\\[8mm]
\textnormal{\normalsize CERN-SPSC-2013-024~/~SPSC-EOI-010}\\%[-3mm]
%\textnormal{~}\\
\textnormal{\normalsize {\today}}\\
%\textnormal{\normalsize version 11.1}\\[4mm]
\textnormal{\normalsize ~}\\[4mm]
\end{flushright}
\begin{center}
%C{\LARGE Search for heavy neutral leptons}
{\LARGE Proposal to Search for Heavy Neutral Leptons at the SPS}
%C{\LARGE Experiment to Search for Heavy Neutral Leptons at the SPS}
\end{center}
}

\vspace{6mm}

%%%%%%%%%%%%%%%%%%%%%%%%%%%%%%%%%%%%%%%%%%
\begin{flushleft}
\small
W.~Bonivento$^{1,2}$,
A.~Boyarsky$^{3}$,
H.~Dijkstra$^{2}$,
U.~Egede$^{4}$,
M.~Ferro-Luzzi$^{2}$, 
B.~Goddard$^{2}$, 
A.~Golutvin$^{4}$,
D.~Gorbunov$^{5}$,
R.~Jacobsson$^{2}$,
J.~Panman$^{2}$,
M.~Patel$^{4}$,
%F.~Rademakers$^{2}$,
O.~Ruchayskiy$^{6}$,
T.~Ruf$^{2}$,
N.~Serra$^{7}$,
M.~Shaposhnikov$^{6}$,
D.~Treille$^{2\,(\ddagger)}$\\
\bigskip
{\footnotesize \it
$ ^{1}$Sezione INFN di Cagliari, Cagliari, Italy\\
$ ^{2}$European Organization for Nuclear Research (CERN), Geneva, Switzerland\\
$ ^{3}$Instituut-Lorentz for Theoretical Physics, Universiteit Leiden, Niels Bohrweg 2, Leiden, The Netherlands\\
$ ^{4}$Imperial College London, London, United Kingdom\\
$ ^{5}$Institute for Nuclear Research of the Russian Academy of Sciences (INR RAN), Moscow, Russia\\
$ ^{6}$Ecole Polytechnique F\'{e}d\'{e}rale de Lausanne (EPFL), Lausanne, Switzerland\\
$ ^{7}$Physik-Institut, Universit\"{a}t Z\"{u}rich, Z\"{u}rich, Switzerland\\
%$ ^{(*)}$spokesperson, Andrey.Golutvin@cern.ch \\
%$ ^{(\dagger)}$local contact, Richard.Jacobsson@cern.ch \\
$ ^{(\ddagger)}$retired \\
%$ ^{8}$Eidgenoessische Tech. Hochschule Z\"{u}rich (ETH Z\"{u}rich), Z\"{u}rich, Switzerland\\
{~}\\
{~}\\
{~}\\
}
\end{flushleft}
%%%%%%%%%%%%%%%%%%%%%%%%%%%%%%%%%%%%%%%%%%
%\date{}
%\maketitle

\vspace{4mm}

% comment this line to turn them off
%\linenumbers

\begin{abstract}

A new fixed-target experiment at the CERN SPS accelerator is proposed that will use decays of charm mesons to search for Heavy Neutral Leptons (HNLs), which are right-handed partners of the Standard Model neutrinos. 
The existence of such particles is strongly motivated by theory, as they can simultaneously explain 
the baryon asymmetry of the Universe, account for the pattern of neutrino masses and oscillations and provide a Dark Matter candidate.

Cosmological constraints on the properties of HNLs now indicate that the majority of the interesting parameter 
space for such particles was
% such particles were mostly
beyond the reach of the previous searches at the PS191, BEBC, CHARM,
 CCFR and NuTeV experiments. 
For HNLs with mass below $2\,$GeV, the proposed experiment will improve on the 
sensitivity of previous searches by four orders of magnitude and will cover a 
major fraction of the parameter space favoured by theoretical models.

%To produce a large number of charm mesons 
The experiment requires a 400\,GeV proton beam from the SPS with a total of $2\times10^{20}$
%C protons on target, achievable in a few years of data taking. 
protons on target, achievable within five years of data taking. 
%C The proposed detector will reconstruct exclusive HNL decays and, in
%C contrast to previous experiments, will measure the HNL mass.
The proposed detector will reconstruct exclusive HNL decays and measure the HNL mass.
The apparatus is based on existing technologies and consists of a target, a hadron absorber, a muon shield, 
a decay volume and two magnetic spectrometers, each of which has a 0.5\,Tm magnet, a calorimeter and a muon detector. 
The detector has a total length of about 100\,m with a 5\,m diameter. 
The complete experimental set-up could be accommodated in CERN's North Area. 

The discovery of a HNL would have a great impact on our understanding of nature and open a new area for future research. 
\end{abstract}

%\linenumbers
%%%%%%%%%%%%%%%%%%%%%%%%%%%%%%%%%%%%%%%%%%%%%%%%%
%\input{introduction}
\section{Introduction}
\label{sec:intr}

The new scalar particle with mass $M_H=125.5\pm0.2_{stat}~^{+0.5}_{-0.6}$$_{syst}$
GeV (ATLAS)~\cite{Aad:2012tfa}, $M_H=125.7\pm0.3_{stat}\pm0.3_{syst}$ GeV (CMS)~\cite{Chatrchyan:2012}, recently
found at the LHC, has properties consistent with those of the
long-awaited Higgs boson of the Standard Model (SM)~\cite{cerutti}. 
This discovery implies that the Landau pole in the Higgs self-interaction
is well above the quantum gravity scale $M_{Pl} \simeq 10^{19}$ GeV (see,
e.g.\ Ref.~\cite{Ellis:2009tp}).   Moreover, within the SM, the vacuum is
stable, or metastable with a lifetime exceeding that of the Universe
by many orders of magnitude~\cite{Buttazzo:2013uya}.
Without the addition of any further new particles, the SM is therefore an 
entirely self-consistent, weakly-coupled, effective field
theory all the way up to the Planck scale (see Refs.~\cite{Buttazzo:2013uya,Bezrukov:2012sa} for a recent
discussion).

Nevertheless, it is clear that the SM is incomplete. 
Besides a number of fine-tuning problems (such as the hierarchy and strong CP problems), the SM is in
conflict with the observations of non-zero neutrino
masses, the excess of matter over antimatter in the Universe, and the
presence of non-baryonic dark matter. 

The most economical theory that can account simultaneously for
neutrino masses and oscillations, baryogenesis, and dark matter, is the
neutrino minimal Standard Model ($\nu$MSM)~\cite{Asaka:2005an,Asaka:2005pn}. 
It predicts the existence of three Heavy Neutral Leptons (HNL) 
and provides a guideline for the required experimental sensitivity~\cite{Gorbunov:2007ak}. 
The search for these HNLs is the focus of the present proposal. 

\newcommand{\ellp}{l^+}
\newcommand{\ellm}{l^-}
In addition to HNLs, the experiment will be sensitive to many other
types of physics models that produce weakly interacting exotic
particles with a subsequent decay inside the detector volume, see e.g.\ 
Refs.~\cite{Batell:2009jf,Gorbunov:2000th,Dedes:2001zia,Faraggi:1999bm,Gninenko:2009ks,Aliev:2007gr}.
%In one generic class of these models~\cite{Batell:2009jf}, the exotic
%particle is created in flavour changing neutral-current processes,
%$D\to \pi X$, followed by a decay to a pair of leptons,
%$X\to\ellp\ellm$. 
%Examples of such models are those with light massive paraphotons~\cite{Batell:2009jf}, 
%supersymmetric models with light sgoldstinos~\cite{Gorbunov:2000th} and 
%R-parity violating neutralinos~\cite{Dedes:2001zia}, 
%bulk singlet neutrinos in models with extra-dimensions~\cite{Faraggi:1999bm}, 
%models with singlet neutrinos with dipole transition moments \cite{Gninenko:2009ks}, 
%and models with unparticles~\cite{Aliev:2007gr}.%
%Examples of such models  are
%$R$-parity violating neutralinos~\cite{Dedes:2001zia}, bulk singlet
%neutrinos in extra-dimension models~\cite{Faraggi:1999bm} and
%Unparticles~\cite{Aliev:2007gr}. 
%In models with non-universal couplings
%to the quarks, the production will not suffer from GIM suppression and
%the sensitivity (if $m_X < m_D$, where $m_D$ is the $D$-meson mass) will be much better than that obtained from 
%$B \to K \ellp \ellm$ decays.%
Longer lifetimes and smaller couplings would be accessible compared to analogous searches
performed previously by the CHARM experiment~\cite{Bergsma:1985qz}.

In the remainder of this document the theoretical motivation for HNL
searches is presented in Section~\ref{sec:theo} and the limits from
previous experimental searches are then detailed in
Section~\ref{sec:exp_status}. The  proposed experimental set-up is 
presented in Section~\ref{sec:exp} and in Section~\ref{sec:bg} the
background sources are discussed, before the expected sensitivity is
calculated in Section~\ref{sec:sens}. 
The conclusions are presented in Section~\ref{sec:conc}.

\section{Theoretical motivation}
\label{sec:theo}

In type-I seesaw models (for a review see Ref.~\cite{Mohapatra:1998rq}) the
extension of the SM fermion sector by three right-handed (Majorana) leptons, $N_I$, where
$I=(1,2,3)$, makes the leptonic sector similar to the quark sector
(see Fig.~\ref{ferm}). Irrespective of their masses, these neutral
leptons can explain the flavour oscillations of the active neutrinos. 
Four 
%Several
different domains of HNL mass, $M_N$, are usually considered:

%%%%%%%%%%%%%%%%%%%%%%%%%%%%%%%%%%%%%%%%%%%%%%%%%
\begin{figure}[!b]
\centerline{
\includegraphics[width=0.48\textwidth]
{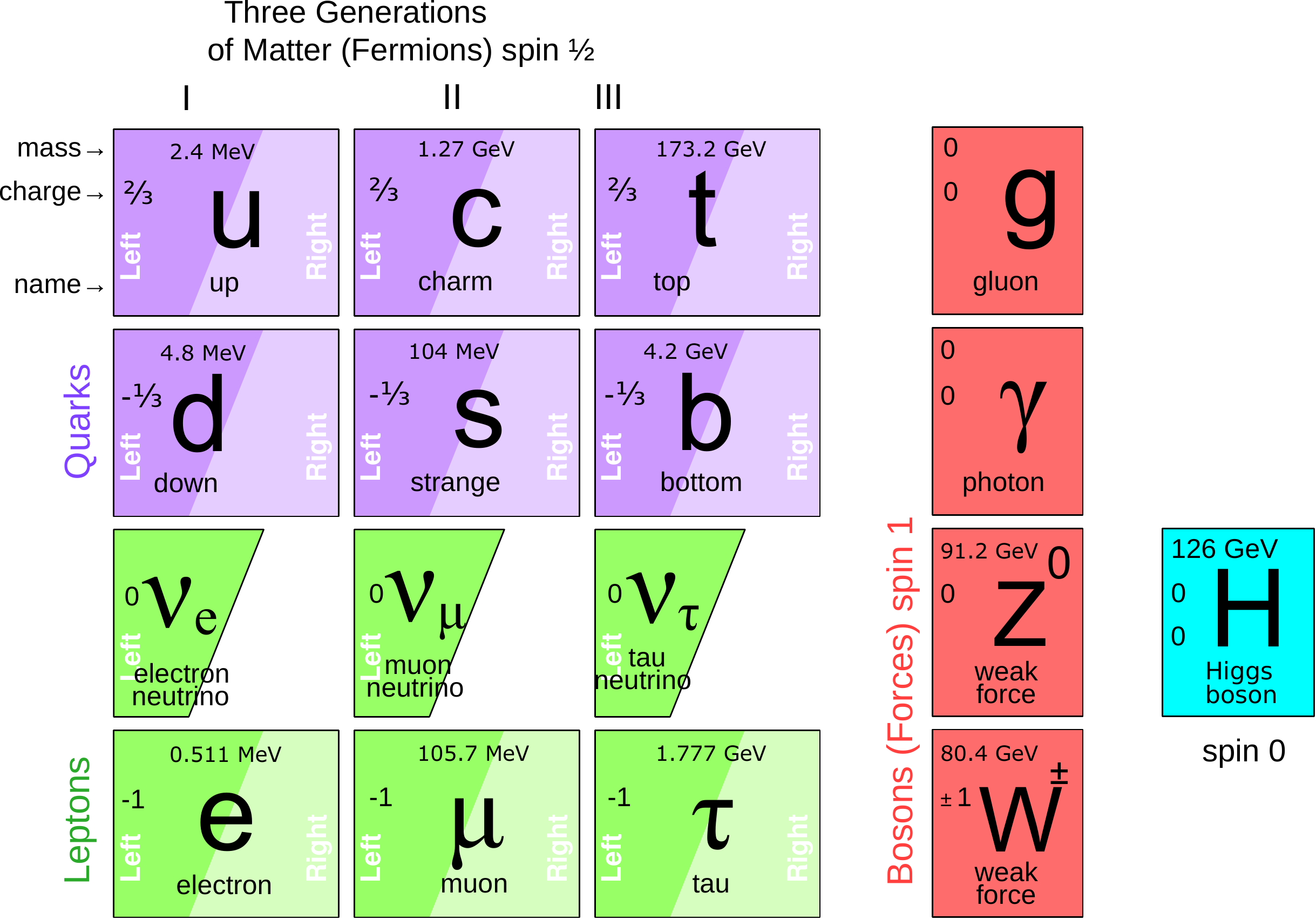}
\hfill
\includegraphics[width=0.48\textwidth]
{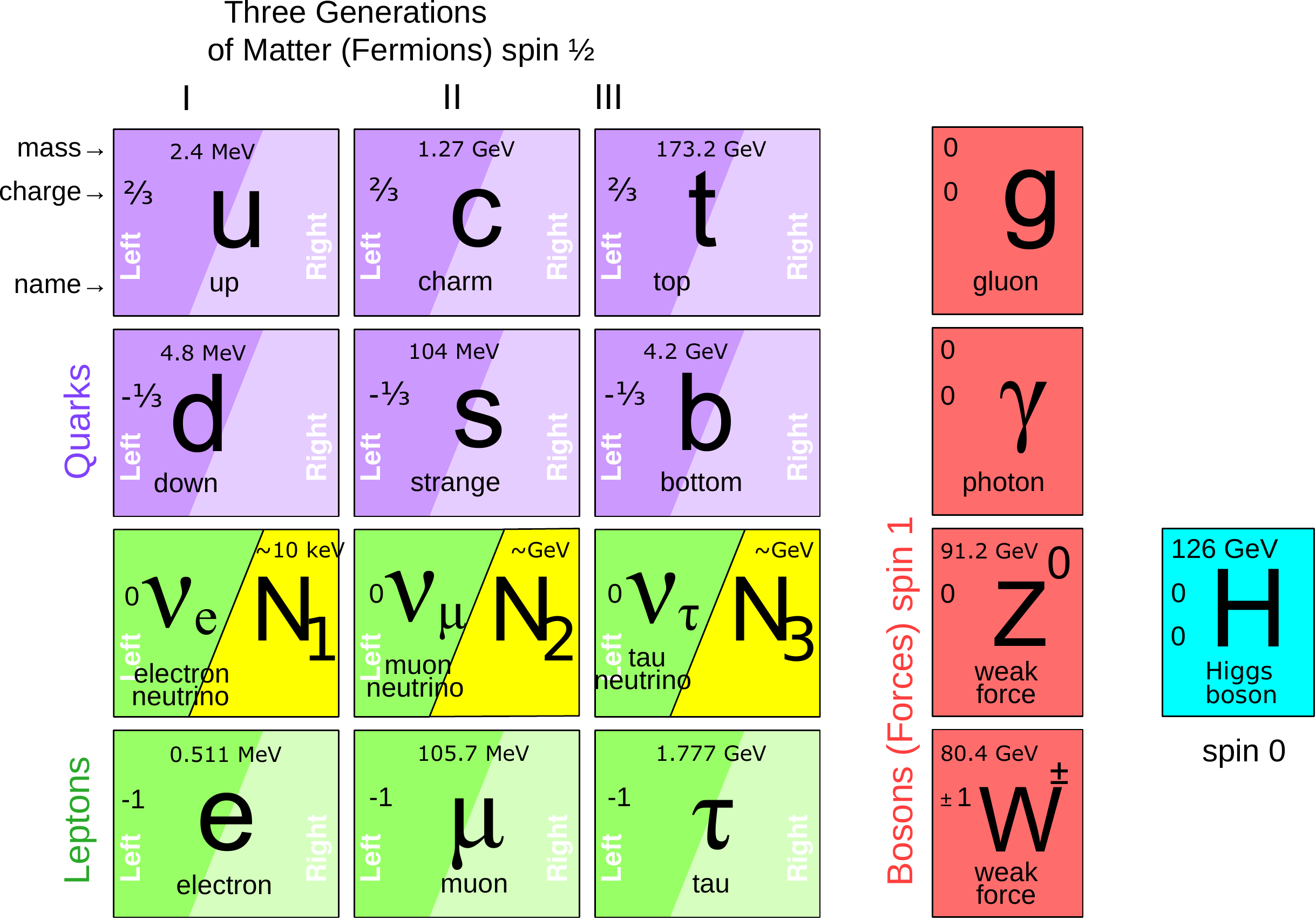}
}
\caption{\small Particle content of the SM and its minimal extension in the
neutrino sector. In the (left) SM the right-handed partners of
neutrinos are absent. In the (right) $\nu$MSM all fermions have both
left- and right-handed components and masses below the Fermi scale.}
\label{ferm}
\end{figure}
%%%%%%%%%%%%%%%%%%%%%%%%%%%%%%%%%%%%%%%%%%%%%%%%%

\begin{itemize}

\item[(1)]{Models with HNLs with $10^9 < M_N < 10^{14}$\,GeV~\cite{Seesaw} 
are motivated by Grand Unified Theories. In such
theories the observed baryon asymmetry of the Universe originates in
CP-violating decays of the HNLs, which produce a lepton
asymmetry~\cite{Fukugita:1986hr}. This asymmetry is then converted into
a baryon asymmetry by sphalerons~\cite{Kuzmin:1985mm,Klinkhamer:1984di}.
The large mass of the HNLs results in a fine-tuning problem for the
Higgs mass. A natural solution is provided by low energy supersymmetry
but at present this is not supported by experimental evidence.
Theories with very heavy neutral leptons
% HNLs 
are unable to account for dark matter
and cannot be directly probed by experiments;}

\item[(2)]{Models with $M_N \sim 10^2-10^3$ GeV (for a review see
Ref.~\cite{Senjanovic:2011zz}) are motivated by a possible solution to the
hierarchy problem at the electroweak scale~(see
e.g.\ Ref.~\cite{Giudice:2008bi}).  The baryon asymmetry of the Universe can
be produced via resonant leptogenesis and
sphalerons~\cite{Pilaftsis:2003gt}. As above, there is no candidate
for dark matter particles. A portion of the parameter space
can be accessed by direct searches at the ATLAS and CMS
experiments~\cite{Pilaftsis:2005rv};}

\item[(3)]{Models with masses of the HNLs below the Fermi scale and
roughly of the order of the masses of the known quarks and leptons,
are able to account for neutrino masses and oscillations and can also give
rise to the  baryon asymmetry of the Universe and can provide dark
matter\cite{Asaka:2005an,Asaka:2005pn,Akhmedov:1998qx,Dodelson:1993je,Shi:1998km} 
(for a review see Ref.~\cite{Boyarsky:2009ix}).
% -- the keV-scale HNL~\cite{Asaka:2005an,Asaka:2005pn,Akhmedov:1998qx,Dodelson:1993je,Shi:1998km}, . 
The phenomenology of GeV-scale HNLs was previously studied in 
Refs.~\cite{Shrock:1980ct,Shrock:1981wq,Gronau:1984ct,Johnson:1997cj}. 
Owing to its relatively large mass, the dark matter candidate -- 
the ${\cal O}(10)$\,keV HNL, does not
contribute to the number of relativistic neutrino species measured recently by
the Planck satellite~\cite{Ade:2013zuv}};

\item[(4)]{Models with $M_N \sim$~eV~\cite{deGouvea:2005er} are
motivated by the $2$--$3\sigma$ deviations observed in short-baseline
neutrino-oscillation experiments~\cite{Aguilar:2001ty,Aguilar-Arevalo:2013pmq}, 
reactor neutrino experiments~\cite{Mention:2011rk} and gallium solar 
neutrino experiments~\cite{Abdurashitov:2005tb,Abdurashitov:1998ne,Hampel:1997fc,Kaether:2010ag}.
Such neutral leptons are usually referred to as sterile neutrinos.
Theories involving these sterile neutrinos can explain neither the
baryon asymmetry of the Universe nor dark matter.}

\end{itemize}
The GeV-scale HNLs of category (3) are able to solve all major problems of
the SM and the search for such particles is the focus of the present proposal.

\subsection{The Neutrino Minimal Standard Model}
\label{ssec:nmsm}

The most general
renormalisable Lagrangian of all SM particles and three singlet
(with respect to the SM gauge group) fermions, $N_I$, is  
\begin{equation}
L_{\rm singlet}=i\bar N_I \d_\mu\gamma^\mu N_I - Y_{I\alpha}\bar N^c_I
\tilde H L^c_\alpha
- M_I \bar {N_I}^c N_I + \rm{h.c.},
\label{pmm}
\end{equation}
where $L_\alpha$, $\alpha=e,\mu,\tau$ are the SM lepton doublets, $c$
is the superscript denoting the charge conjugation,
$\tilde{H}_i=\epsilon_{ij}H_j^*$, where $H$ is the SM Higgs doublet,
and $Y_{I\alpha}$ are the relevant Yukawa couplings. The last term is
the Majorana mass term, which is allowed as the $N_I$ carry no gauge
charges.  When the SM Higgs field gains a non-zero vacuum expectation
value, $v=246$\,GeV, the Yukawa term in Eqn.~\eqref{pmm} results in
mixing between the HNLs and the SM neutrinos. In the mass basis, the 
massive active-neutrino states mix with each other, as is required to
explain neutrino oscillations.

The model given by Eqn.~\eqref{pmm} contains eighteen new parameters compared to the SM 
(six CP-violating phases, six mixing angles, three Dirac masses, and
three Majorana masses).  
Five of these parameters (related to three mixing angles of the active neutrinos and two mass differences) 
have been determined by low-energy neutrino experiments~\cite{gonzalez:2012}.
The large number of CP-violating phases opens
the possibility of significantly larger CP violation than that seen in
the quark sector.  
In particular, the baryon asymmetry of the Universe can be explained by a wide range of parameters~\cite{Drewes:2012ma}.
In fact, HNLs with any mass difference and mixing angle allowed by experimental constraints can produce the 
necessary baryon asymmetry in the Universe (BAU).

The most interesting variant of this model is the $\nu$MSM. 
In this model, the lightest singlet fermion, $N_1$, has a very weak mixing with the other leptons, 
playing no role in active-neutrino mass generation. 
The lightest singlet $N_1$ is then sufficiently stable to be a dark matter candidate. 
This particle could be detected by searching for a narrow line in the X-ray
spectrum coming from radiative decays $N_1\to\nu\gamma$ (for a review
see Ref.~\cite{Boyarsky:2009ix}).  
The reduced number of parameters that arises from requiring that $N_1$ be a dark matter candidate results in the further requirement that
$N_{2,3}$ be nearly degenerate in mass.
This enables CP violation to be enhanced to the level required to explain the baryon asymmetry in the Universe.

The order of magnitude of the different parameters in Eqn.~\eqref{pmm}
can be understood by considering the typical value of the Dirac mass
term, $m_D \sim Y_{I\alpha} v$. The scale of the active neutrino
masses is given by the seesaw formula,
\begin{equation} 
m_{\nu} \sim \frac{m_D^2}{M}\,.
\end{equation}
\noindent For the HNL mass $M\sim 1$\,GeV, and $m_{\nu}\sim 0.05$\,eV, the
value of $m_D$ is in the region of $10$\,keV, and the Yukawa couplings
are of the order of $10^{-7}$.

Denoting the mixing angles between $N_I$ and active neutrinos of
flavour $\alpha$ by $U_{I\alpha} = Y_{I\alpha} v/\sqrt{2}M$, the
integral mixing angle,
\begin{equation} 
U^2 = \sum_{I,\alpha} |U_{I\alpha}|^2,
\end{equation}
\noindent measures the overall strength of interactions between
the $N_I$  and the active neutrinos. At a given mass, $M$, a larger
$U$ yields stronger interactions (mixing) of the singlet fermions and
the SM leptons. In the early Universe, large $U$ would result in the
$N_{2,3}$ particles coming into equilibrium above the electroweak
temperature and would therefore erase any baryon asymmetry in the Universe. 
A small
mixing between the neutral leptons and the active neutrinos would
enable HNLs to generate the observed baryon asymmetry of the Universe and would also
explain why these particles have not yet been observed in experiments.

\subsection{Heavy Neutral Leptons -- production and decay mechanisms}
\label{ssec:prod_decay}

%%%%%%%%%%%%%%%%%%%%%%%%%%%%%%%%%%%%%%%%%%%%%%%%%%
\begin{figure}[!tb]
  \vspace*{-20mm}
  \centerline{
    %plots from 1112.3319v1
    \includegraphics[width=0.75\textwidth]{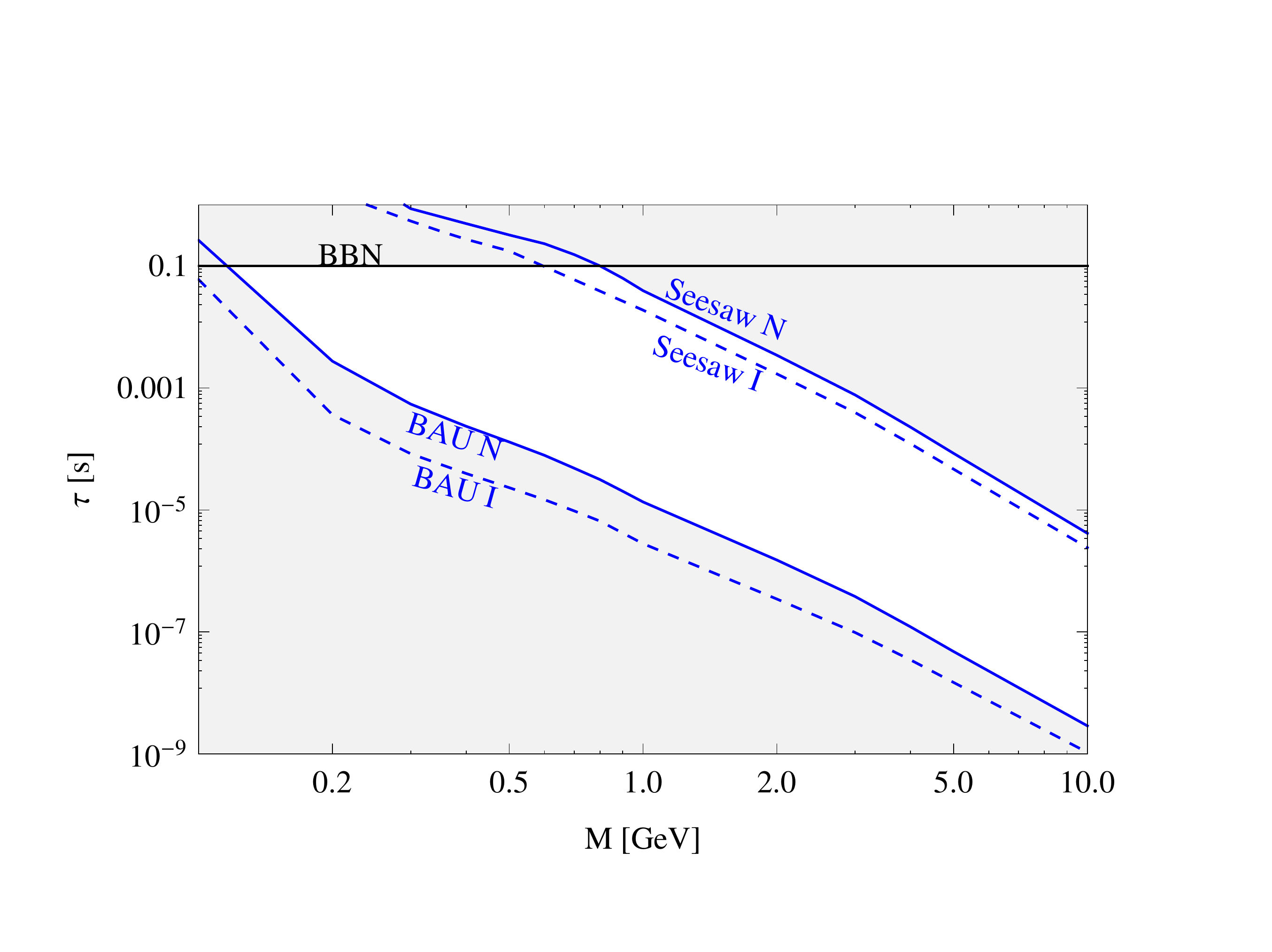}
  }
  \vspace*{-10mm}
 \caption{\small Constraints on the HNL lifetime, $\tau$, from Big Bang Nucleosynthesis
 (black line: ``BBN''), from the
 baryon asymmetry of the Universe (``BAU'') and from the seesaw mechanism (blue solid lines: ``BAU N'' and ``Seesaw N''
 refer to a normal mass-hierarchy of active neutrinos and ``BAU I'' and ``Seesaw I'' refer to an
 inverted mass-hierarchy). The allowed region of the parameter space is
 shown in white for the normal hierarchy case. 
 The limits from direct experimental searches are outlined in Fig.~\ref{exp}. Figure taken from Ref.~\cite{Canetti:2012vf}.}
 \label{TauN}
\end{figure} 
%%%%%%%%%%%%%%%%%%%%%%%%%%%%%%%%%%%%%%%%%%%%%%%%%%%%%%%%
\begin{figure}[!tb]
\centerline{
\hfill
\includegraphics[width=0.3\textwidth]{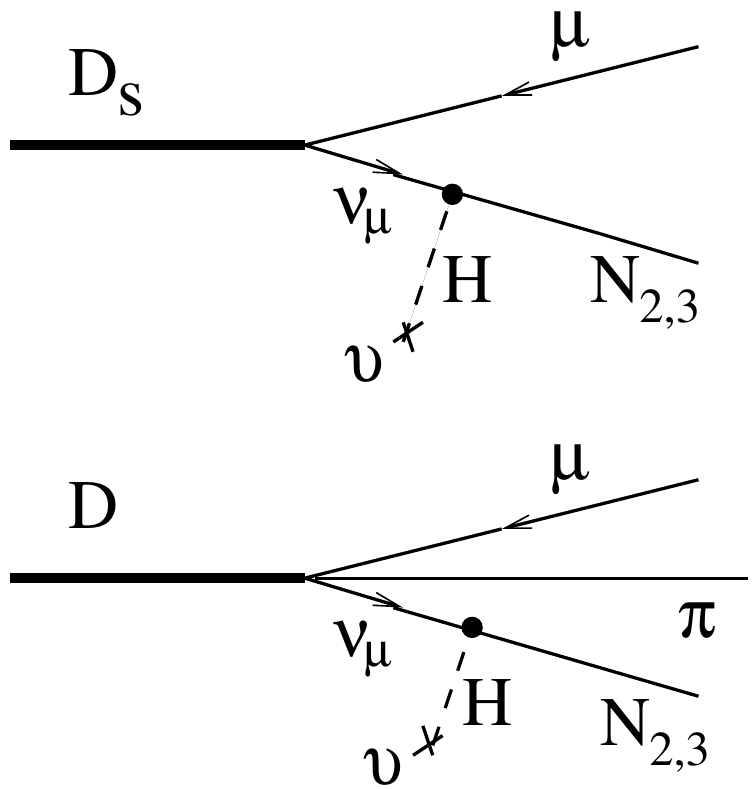}
\hfill
\includegraphics[width=0.3\textwidth]{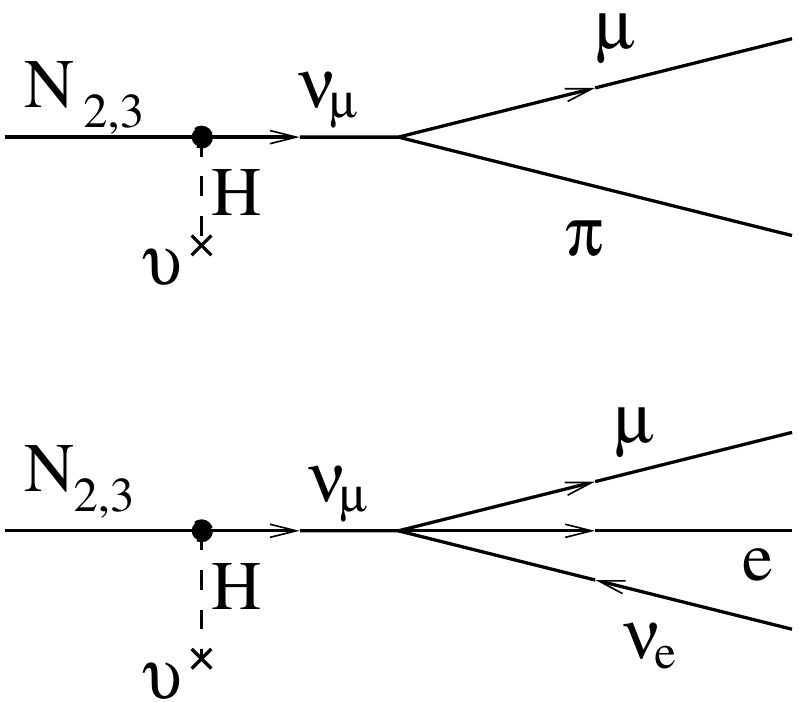}
\hfill
}
\caption{\small Feynman  diagrams (left) for the production of HNLs and (right)
 for their decays. The dashed line denotes the coupling to the Higgs vacuum expectation value,
leading to the mixing of active neutrinos and HNLs via Yukawa couplings.}
\label{production-and-decays}
\end{figure}
%%%%%%%%%%%%%%%%%%%%%%%%%%%%%%%%%%%%%%%%%%%%%%%%%%
%
In the $\nu$MSM the HNL-neutrino mixing gives rise to HNL ($N_{2,3}$)  production in
weak decays of heavy mesons. The same mixing gives rise to the decay of the HNLs
to SM particles. % In the $\nu$MSM
The allowed mixing angles are small and the $N_{2,3}$ particles are
much longer-lived (by a factor $\sim 1/U^2$) than weakly decaying SM
particles of similar mass (see Fig.~\ref{TauN}). 
For HNL masses below the charm threshold, the most relevant production
mechanisms are shown in 
Fig.~\ref{production-and-decays}~(left). The requirement for mixing into
active neutrinos at both production and decay results in 
signal yields which depend on the fourth power of the HNL-neutrino mixing, $U^4$.

Potential two- and three-body decay modes of $N_{2,3}$ are shown in
Fig.~\ref{production-and-decays}~(right).
For $D$ mesons, the typical branching fractions
expected for the upper and lower limits of the $\nu$MSM parameter
space  are at the level of~\cite{Gorbunov:2007ak}
\begin{equation}
\label{D-branching}  
{\cal B}_{D\to N}\sim 10^{-8}-10^{-12}. 
\end{equation}

The three-body leptonic decay branching fractions depend on the
flavour pattern of HNL-to-active neutrino mixing:  
${\cal B}(N\to\nu l_1^+l_2^-)\simeq 1-10\%$ ($l_{1,2}=e,\mu$).  
Among the two-body decays the most promising for searches are\footnote{
In the following, charge conjugation is implied unless stated otherwise.}  
${\cal B}(N \to e^-\pi^+)\simeq 0.2\%-50\%$, ${\cal B}(N\to \mu^-\pi^+)\simeq 0.1\%-50\%$, 
${\cal B}(N \to e^-\rho^+)\simeq 0.5\%-20\%$, ${\cal B}(N\to \mu^-\rho^+)\simeq 0.5\%-20\%$.  
Branching fractions to $e^-K^+$ and $\mu^-K^+$ are always below 2\%.  
The $\mu^-\pi^+$ final state is  the cleanest signature experimentally and is the
focus of the studies below. The $\mu^- \rho^+$ and $e^-\pi^+$ final states 
provide additional experimental signatures that extend the sensitivity and could be used to constrain additional parameter space. 

Assuming a branching fraction $10^{-12}$ and a factor $10^{-4}$ from the lifetime,
an experiment to detect
$N_{2,3}$ would require more than $10^{16}$ 
$D$ mesons in order to fully explore the parameter space with $M < 2$\,GeV. 
Preliminary studies of an experimental design were described in Ref.~\cite{original-proposal}.

\section{Experimental status and cosmological constraints}
\label{sec:exp_status}
\begin{figure}[!tb]
\centerline{
\includegraphics[width=0.48\textwidth]{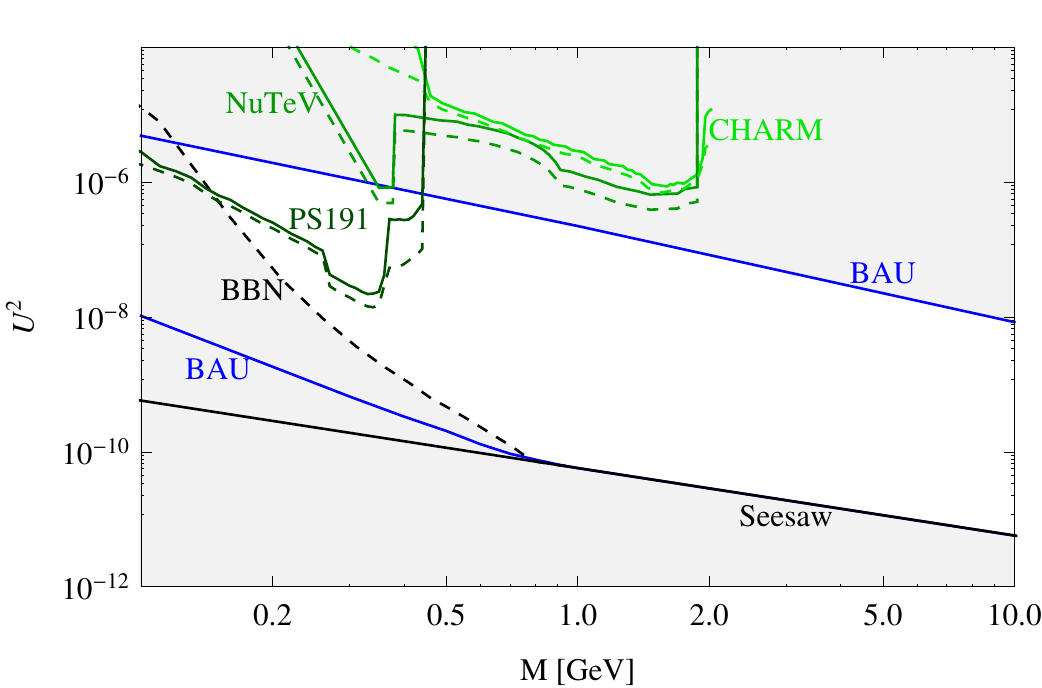}\hskip
0.04\textwidth
\includegraphics[width=0.48\textwidth]{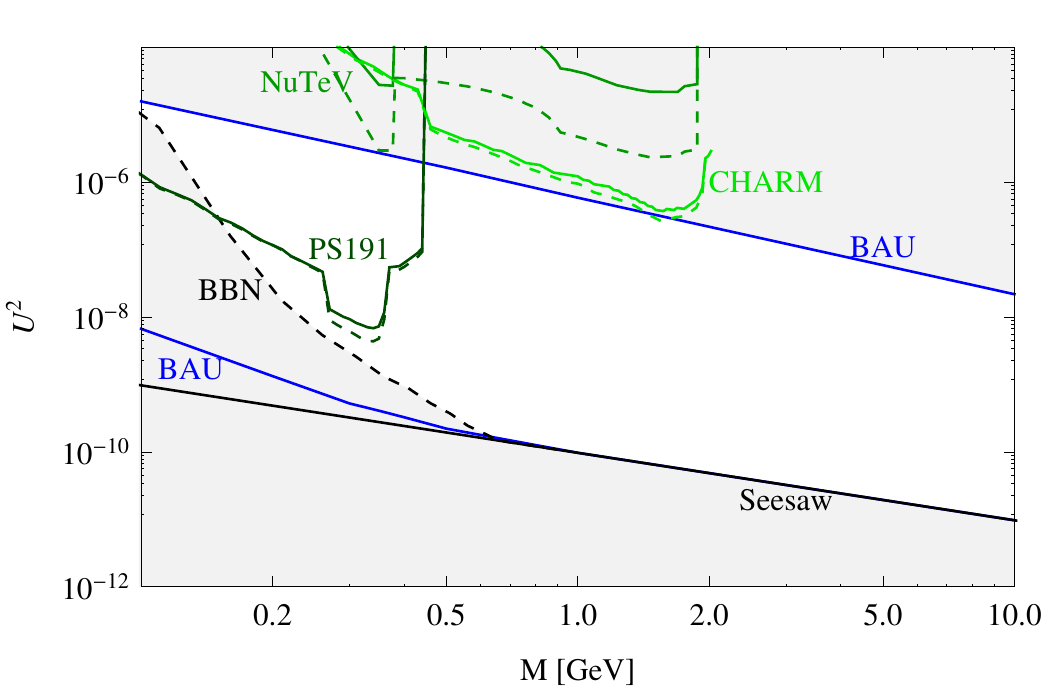}}
\caption{\small
 Current bounds on the parameters of the degenerate HNLs, $N_{2,3}$ in
 the ($U^2$, $M_N$) plane from 
 cosmological considerations and neutrino mixing together with limits
 from previous experimental searches (the solid
 and dashed lines indicate the dependence of these regions on the pattern
 of HNL mixing with the electron, muon and tau-neutrino).
 Figures taken from Ref.~\cite{Canetti:2012vf}.
 A normal mass hierarchy of the neutrinos is shown on the left and an
 inverted hierarchy on the right.}
\label{exp}
\end{figure}

The region of the lifetime-mass $(\tau,\,M_N)$ plane consistent with the
cosmological constraints is shown in Fig.~\ref{TauN}.  
Figure~\ref{exp} shows the allowed region in the $(U^2,\,M_N)$ plane,
given the  constraints  from particle physics experiments. 
For all points in Fig.~\ref{exp} below the line marked ``Seesaw'', 
 the mixing of the HNL with active neutrinos becomes too weak to produce the observed pattern of neutrino
flavour oscillations. 
Cosmological considerations result in additional limits. 
If the HNLs are required to provide a mechanism for baryogenesis, their
coupling with matter should be sufficiently weak such that they lie
below the upper line marked ``BAU''.  
A HNL with the parameters to the left of the ``BBN'' line would live
sufficiently long in the early Universe to result in an overproduction
of primordial Helium-4 in Big Bang
Nucleosynthesis~\cite{Dolgov:2000jw,Ruchayskiy:2012si}.  
The regions excluded by the CHARM~\cite{Bergsma:1985is}, CERN
PS191~\cite{Bernardi:1985ny} and NuTeV~\cite{Vaitaitis:1999wq}
experiments are also shown.  
Limits by BEBC~\cite{CooperSarkar:1985nh} and CCFR~\cite{ccfr:1987} are not shown in Fig.~\ref{exp}.
%The region excluded by CCFR in similar to that by BEBC.
A detailed discussion of the experimental constraints (including also
those from peak search experiments~\cite{peaksearch}) 
% \cite{Shrock:1980vy,Shrock:1980ct,Shrock:1981wq,Shrock:1981cq}, \cite{Aoki:2011vma,Britton:1992pg,Britton:1992xv,Yamazaki:1984sj, Bryman:1996xd,Abela:1981nf,Daum:1987bg,Hayano:1982wu} 
 is presented in Refs.~\cite{Gorbunov:2007ak,Atre:2009rg,Ruchayskiy:2011aa}.

%%%%%%%%%%%%%%%%%%%%%%%%%%%%%%%%%%%%%%%%%%%%%%%%%
%%%%%%%%%%%%%%%%%%%%%%%%%%%%%%%%%%%%%%%%%%%%%%%%%

The combined experimental and theoretical constraints imply that the HNLs must have masses larger than approximately $100$\,MeV. 
%\marginpar{check}
The domain of masses accessible by the present experiment, i.e.\ masses below those of charm mesons,
naturally appears under the assumption that the observed hierarchy in the
masses of the different generations of quarks and charged leptons is
preserved in the Majorana sector of the theory. 

\section{Experimental set-up}
\label{sec:exp}

\begin{figure}[tb]
  \vspace*{-2mm}
  \begin{center}
  \includegraphics[width=0.49\textwidth]{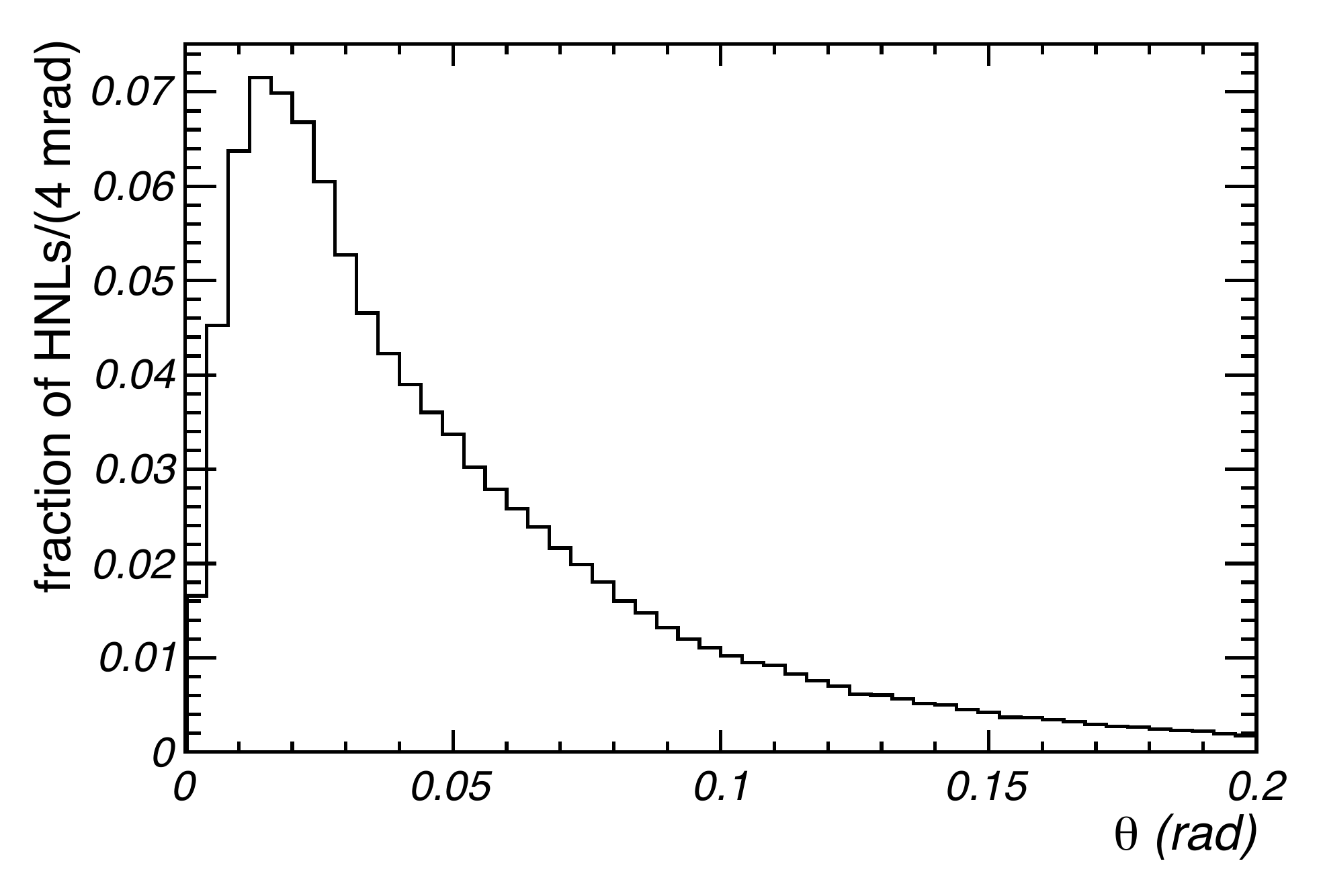}
  \hfill
  \includegraphics[width=0.49\textwidth]{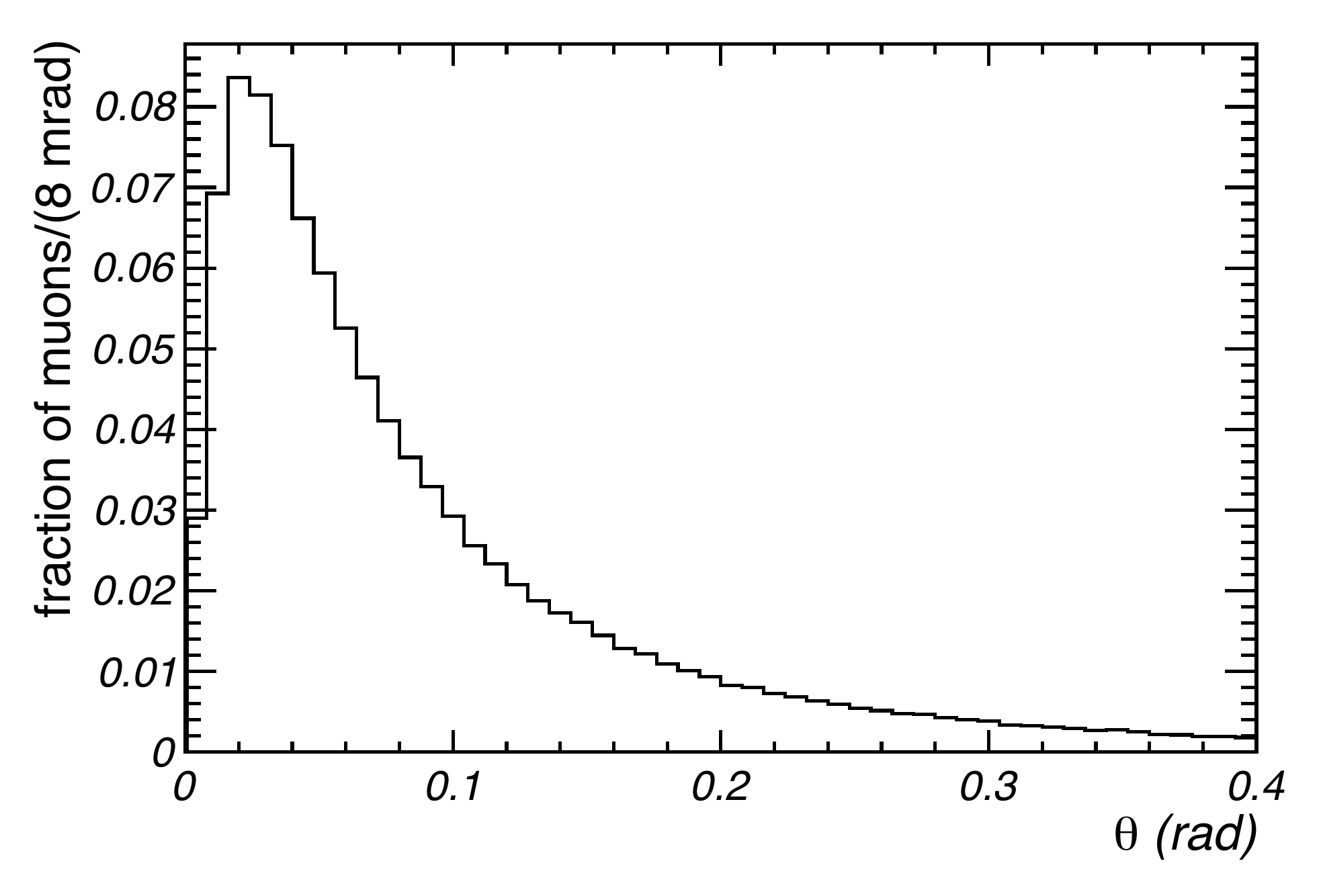}
  \end{center}
  \vspace*{-6mm}
  \caption{\small Polar angle distribution (left) of the HNLs and (right) of the
    muons and pions from the decay $N \rightarrow\mu^{-}\pi^{+}$ in simulated HNL decays with $M_N = 1$~GeV.}
  \label{pthnl}
\end{figure}

The proposed experiment will use a 400\,GeV proton beam on a fixed target
to produce a large number of charm mesons. 
The HNLs from charm meson decays have a significant polar angle with respect to the beam direction, 
$\approx 50$\,mrad on average, as shown in Fig.~\ref{pthnl}. 
In order to maximise the geometric acceptance for a given transverse
size of the detector, the detection volume must therefore be placed as close as possible to the target.

The production of the charm mesons is accompanied by copious direct production of
pions, kaons and short-lived light resonances. The subsequent decays of these particles would 
result in a large flux of muons and neutrinos. 
To minimise these decays, a combination of a target and a hadron absorber of a few metres length, 
both made of as dense a material as possible, is required. 
To reduce the detector occupancy and
backgrounds induced by the residual muon flux,  a muon 
shield is required downstream of the hadron absorber.
The experimental set-up must therefore balance the opposing requirements of locating the detector 
as close as possible to the target and of accommodating a sufficiently long muon shield 
upstream of the fiducial volume of the detector to reduce muon-induced backgrounds.

The detector  must be able to reconstruct the final state 
particles from $N \to \mu^{-}\pi^{+}$ decays,\footnote{Henceforth, the notation $N$ will be used to indicate $N_{2,3}$.} 
identify muons, and determine the $\mu^{-}\pi^{+}$ 
invariant mass and parent particle flight direction with sufficient resolution to reject backgrounds.
To be sensitive to the decays 
$N \to e^- \pi^+$, $N \to e^- \rho^+$ and $N\to \mu^- \rho^+$ in addition to the $N\to \mu^- \pi^+$ decay, 
a magnetic spectrometer, an electromagnetic calorimeter and a muon detector are mandatory.

The background neutrino flux and the residual muon flux in the detector constitute
crucial parameters in the design of the experiment.
Interactions of neutrinos inside the fiducial volume can mimic signal events,
in particular via charged-current interactions of muon-neutrinos.
This motivates evacuating the fiducial volume to a level where such background
events are negligible (see Section~\ref{sec:bg}). 

Interactions of neutrinos and muons in the material near the fiducial volume, 
can produce long-lived $V^0$ mesons, such as neutral kaons, %$K_L$'s and $K_S$'s, 
which can decay in the detector fiducial volume and mimic signal events. 
To suppress neutrino-induced $V^0$ background events from the downstream end of the muon shield,
the neutrino flux from light meson decays must be minimised at the source.
This is achieved by the use of target and hadron absorber
materials with the smallest possible interaction length.
In addition, the muon shield must be sufficiently long to reduce muon-induced $V^0$ backgrounds 
to a level that is comparable to or smaller than the background from neutrinos.
Shielding against cosmic rays is not required and the detector could
therefore be located in an open space.
A description of the proposed beam line and detector design is given below.
%%%%%%%%%%%%%%%%%%%%%%%%%%%%%%%%%%%%%%%%%%%%%%%%%%
%%%%%%%%%%%%%%%%%%%%%%%%%%%%%%%%%%%%%%%%%%%%%%%%%%

\subsection{Beam line design}
\label{ssec:infra}

The SPS beam energy, beam intensity and the flexibility of the time structure offer a perfectly suitable 
production facility for the proposed experiment. The recently completed CERN Neutrinos to Gran Sasso (CNGS) 
programme made use of $400$\,GeV  protons from the SPS over a nominal year with 200 days of operation, 
55\% machine availability, and 60\% of the SPS supercycle. 
This produced a total of $4.5\times 10^{19}$ protons on target per year. 
The experiment described here assumes the same fraction of beam time as CNGS and the current SPS performance. 
In particular, the preliminary investigation of the operational mode assumes minimal modification to the SPS, 
maximises the use of the existing transfer lines and is compatible with the operation of both the LHC and the
North Area fixed target programmes as they are currently implemented.

Increased experimental reach could come as a by-product of a future SPS upgrade from a proton 
spill intensity of about $4.5\times 10^{13}$ protons per pulse to $7.0\times 10^{13}$ protons per pulse. 
The design of the proposed experiment aims at taking advantage of such an upgrade but the experiment does not rely on it.

\subsubsection{Beam extraction and proton target}
\label{sssec:target}

From the general considerations given above, the experimental requirements favour the choice of a relatively long 
extraction to reduce the detector occupancy per unit time, and to allow a simple target design based 
on dense materials. At the same time, the extraction type should not affect significantly the SPS cycle time 
as compared to CNGS, and should respect the constraints on the activation levels in the extraction region 
and the risk of damage to the extraction septum. Several options exist, all of which may be acceptable from 
the point of view of the experiment, but the optimal choice requires further study:

\begin{itemize}
\item{Slow extraction with a spill length of $1.2$\,s.
This would mean a $7.2$\,s SPS spill cycle, and consequently a reduction 
of the number of protons on target by 10\%;}
\item{Fast non-coherent resonant extraction of O(1\,ms);}
\item{Fast extraction similar to that of CNGS of O($10\,\mu$s).}
\end{itemize}

The first two options involve extraction with the SPS RF system OFF and the beams debunched. This produces a 
quasi-continuous spill that is compatible with a continuous detector
readout, as described in Section~\ref{ssec:synergy}.
The first two extraction methods result in a lower detector occupancy.

With lifetimes of $\sim 10^{-12}$\,s, the dominant fraction of $D$ mesons decay before interacting in the target, 
regardless of the target material. 
However, the flux of secondary pions and kaons, which give rise to a substantial muon flux 
from decays in flight, can be greatly reduced by the use of a dense target material with a 
short interaction length. 
For example, the use of tungsten, rather than graphite, reduces the number of decaying
pions and kaons by an order of magnitude. 
A 50~cm long tungsten target would suffice.
However, the beam energy density must then be diluted in order to avoid destructive 
thermal shock waves in the target and to allow effective heat extraction. 
It may be possible to extract the deposited beam energy using water cooling.

Since the neutrinos produced in charm decays, which can mix with HNLs, have in any case a relatively high transverse 
momentum (see Fig.~\ref{pthnl}), there is no requirement to have a small beam spot. 
The beam line design can then be driven purely by the technical requirements and constraints. 
For the same reason, the experiment does not impose stringent constraints on the optical parameters of the extracted beam. 
The  dilution of the beam energy deposition mentioned above may therefore be obtained 
by allowing the transverse size of the beam to increase in a dedicated section of the primary beam line, and by 
using a combination of orthogonally deflecting kicker magnets to produce
a Lissajous sweep across the target, similar to the LHC beam dump. 
Preliminary investigation shows that it should be possible to achieve a beam spot
size of about $5$\,mm on the target and a sweep diameter of several centimetres, with the only constraint 
being a beam divergence less than the natural divergence of the produced HNLs.
Beam losses in the straight section leading to the target must be minimised and monitored to reduce the background neutrino flux.

The use of a slow extraction would simplify the target design.
To prevent damage to the target caused by possible failures in the beam line, the beam extraction 
would need to be interlocked with the magnet currents in the dedicated transfer line.

The proton target will be required to withstand a beam power of $750$\,kW, corresponding to 
$7\times 10^{13}$ protons per $6$\,s at $400$\,GeV. 
Detailed thermo-mechanical studies will be required in order to produce a  technical design.

\subsubsection{Hadron absorber}
\label{sssec:stopper}

To stop the remaining secondary pions and kaons before they decay, a hadron absorber will immediately follow the target. By
surrounding the target, the absorber will also stop pions and kaons at large angles which may otherwise produce muons
that enter the fiducial volume due to large angle scattering. 
The absorber will also provide the first level of lateral radiation shielding, as well as absorbing the residual non-interacting 
protons ($\sim1\%$ of the incident proton flux) and the electromagnetic radiation generated in the target. 
The physical dimensions of the absorber are driven by the radiological requirements on the muon shield (see Section~\ref{ssec:mushield} below).  
As part of the absorber, a concrete shielding wall will close-off the target bunker volume from the downstream muon shield tunnel.

\subsubsection{Muon shield}
\label{ssec:mushield}

The muon flux after the hadron absorber is estimated from a sample of $10^9$ $pp$ events generated with PYTHIA
with a proton beam energy of 400\,GeV and a fixed proton target.
The prompt component of the muon flux originates from the electromagnetic decays of meson
 resonances, mainly $\eta$, $\rho$, $\phi$, $\omega$.
The non-prompt component originates mainly from the decay in flight of charged pions and kaons. 
For the present estimates, only the decays of primary pions and kaons are considered as non-prompt sources. 
Secondary, tertiary, or higher order pion or kaon decays should give a softer momentum spectrum 
that is more easily attenuated than the high-energy primary pion and kaon decays that drive the 
requirement on the muon shield length.
The exact radial extent of the muon shield will be defined based on a detailed simulation
to ensure that the muons from higher order interactions, 
which generally have larger polar angles, are also absorbed.

\begin{figure}[tb]
   \vspace*{-1mm}
   \begin{center}
   \includegraphics[width=0.7\textwidth]{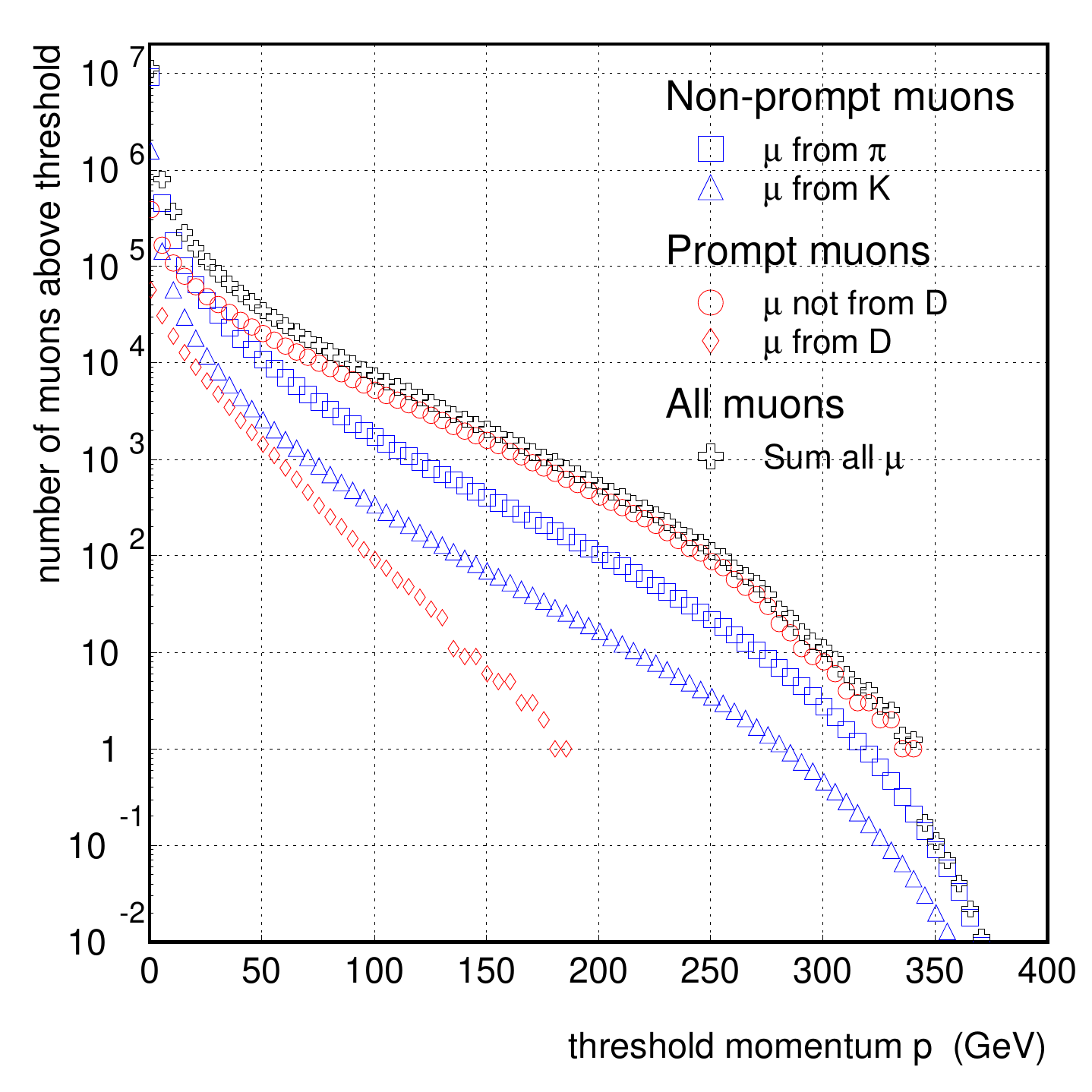}
   \end{center}
   \vspace*{-8mm}
   \caption{\small Estimated muon fluxes after a hadron absorber  based on $10^9$ $pp$ events  
     generated with PYTHIA with a proton beam of 400~GeV and a fixed proton target (see text for details).
     The total flux of first-generation muons above a given momentum is shown. 
     %Left: fluxes per momentum bin (1 GeV/$c$). Right: fluxes above a threshold momentum.  with $\lambda_{\rm int}=11\,$cm
    }
   \label{fig:muonfluxes}
   \end{figure}

The dense target and the hadron absorber described above will be designed to stop all hadrons.
However, a fraction ${\cal B}_\mu \, \lambda_{\rm int}/\lambda_{\rm dec}$ of the hadrons
entering the absorber will decay into muons before interacting, where $\lambda_{\rm int}$  
is the interaction length of hadrons in the hadron absorber;
$\lambda_{\rm dec} = \tau \, p/M$ is the decay length of the hadron and
$\tau$, $p$ and $M$ are the hadron lifetime, lab momentum and rest mass; and
${\cal B}_\mu $ is the branching fraction for the decay of the hadron to muons 
(100\% for $\pi^+ \rightarrow \mu^+ \nu_\mu$ and 63\% for $K^+\rightarrow \mu^+ \nu_\mu$, 
other muon-producing decay channels are neglected).
A cross-check with the muon fluxes observed in the CHARM ``beam dump'' experiment~\cite{Charm-fluxes} 
indicates that the above  method  for estimating the  ``first-generation'' muon flux gives
the correct order of magnitude.
The resulting muon flux is shown as a function of the muon momentum in Fig.~\ref{fig:muonfluxes}.
The muons from charm decays represent a negligible fraction of the total muon flux.

Figure~\ref{fig:muonfluxes} indicates that if a shield were put in place to stop 
muons below 350\,GeV, the muon flux would be reduced by about seven
orders of magnitude to O(1) muon per $10^9$ protons on target.
However, the muon-induced $V^0$ production rate from nuclei would still be larger than that of 
neutrino interactions that is computed in Section~\ref{sec:bg}. 
Therefore, the experiment will use a muon absorber that nominally stops muons with energies of up to 400\,GeV, 
which requires a length of 52\,m of uranium (or 54\,m of tungsten)~\cite{groom}.

\subsubsection{Potential experimental site}
\label{sssec:sites}

\begin{figure}[tb]
  \begin{center}
    \vspace*{-25mm}
    \includegraphics[width=1\textwidth]{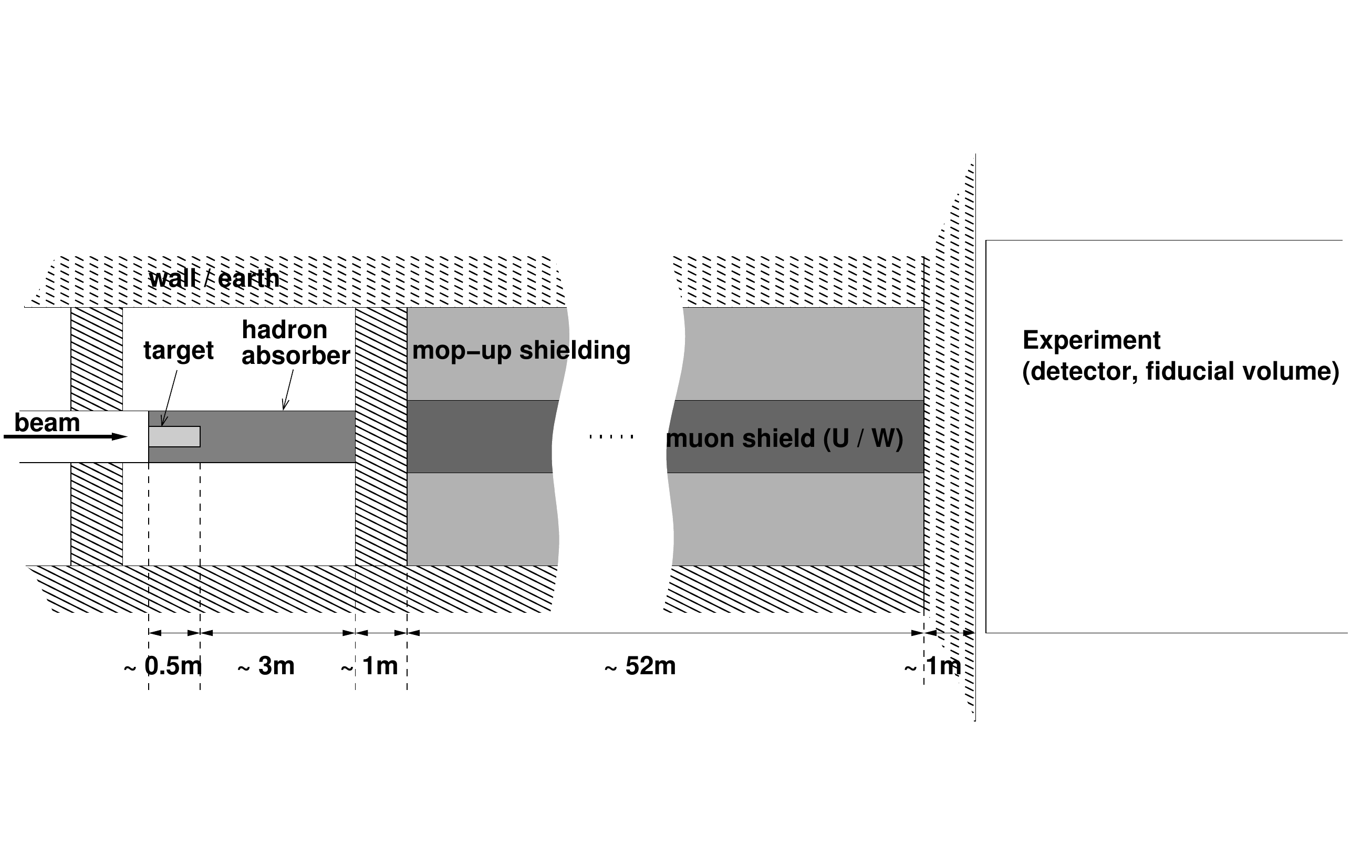}
    \vspace*{-25mm}
  \caption{\small Schematic view of the target, hadron absorber and muon shield in front of the
   experiment. The total length from the target to the entrance of the fiducial volume is $\sim60$\,m.
   }
  \label{fig:setup-sketch}
  \end{center}
  \end{figure}

The final configuration of the beam line and detector will be based on an optimisation of
the proton spill duration, the target design, the hadron absorber and the muon shield lengths, 
and the number of protons on target.

The general beam line layout as described above is shown in Fig.~\ref{fig:setup-sketch}. 
The proposed beam line arrangement consists of a proton beam impinging
on a $\sim 0.5$~m long target, followed by a hadron absorber of $\sim 3$~m length, 
a shielding wall to confine air and radiation, and a $\sim 52$~m long uranium or tungsten
 muon absorber enclosed in iron and concrete.

The experiment requires a new switching section from an existing beam line, which can be very 
similar to the one already used for the CNGS facility. A short section of transfer line, 
consisting almost entirely of drift length to allow the beam energy density to be reduced, will then bring 
the high energy protons to the target bunker that contains the target and the hadron absorber.
The site of the experiment should be such that it allows use of existing 
infrastructure. In particular, the site will need to be provisioned with the 
equipment to handle air and water activation, to comply with radiation protection standards. 
The site should also be well within the CERN boundaries. 
Both construction and operation of the beam line and 
the detector can be such that the interference caused to running 
facilities is minimised.

Based on these requirements and constraints, the North Area could provide a
suitable location with
a relatively short transfer line of a few hundred metres, branching-off from TT20, 
with a new beam splitting near to the TDC2 splitter area. 
The beam deflection could then be arranged to allow the dedicated transfer line
to point to a target bunker at sufficient depth and distance from the TCC2 
target area to minimise the excavation in activated soil. Soil tests will 
be needed to determine the exact location. A 60\,m trenched tunnel after 
the target bunker would allow the installation of the muon shield, and an area
at relatively low elevation just before the North Area hall would 
provide a suitable site for the location of the surface detector building. The 
location of the proposed target bunker near to the TCC2 target area suggests that 
it may be possible to benefit from the general refurbishments that are already 
foreseen for the current target area in terms of activated water and air treatment.

\subsection{Detector}
\label{ssec:detector}

The detector consists of a long decay volume followed by a spectrometer.
For a given detector length, the detector diameter should be maximised.
In the discussion below the 5\,m aperture of the LHCb spectrometer~\cite{Magnet-TDR} is taken as a realistic scale.

\begin{figure}[!tb]
\centerline{
\includegraphics[width=0.55\textwidth]{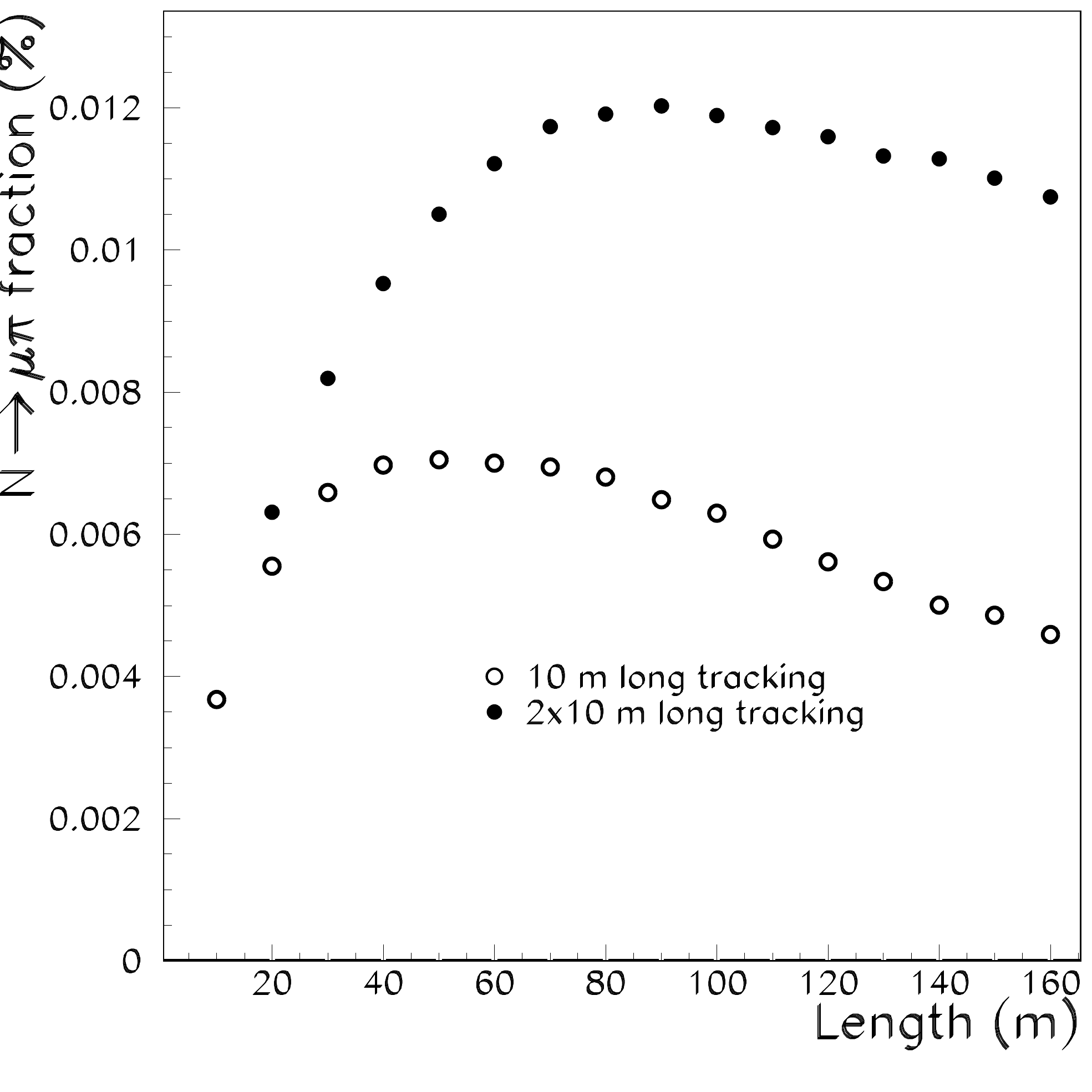}
}
\vspace*{-5mm}
\caption{\small Fraction of HNL in the detector acceptance as a function of the 
length of the fiducial volume. 
Open circles: a single spectrometer following a fiducial volume of a given length.
Full circles: two spectrometers in series, each following a fiducial volume
of half the given length.
The spectrometer length is fixed to 10\,m.
}
\label{Length}
\end{figure}

\begin{figure}[!tb]
\centerline{
  \includegraphics[width=1\textwidth]{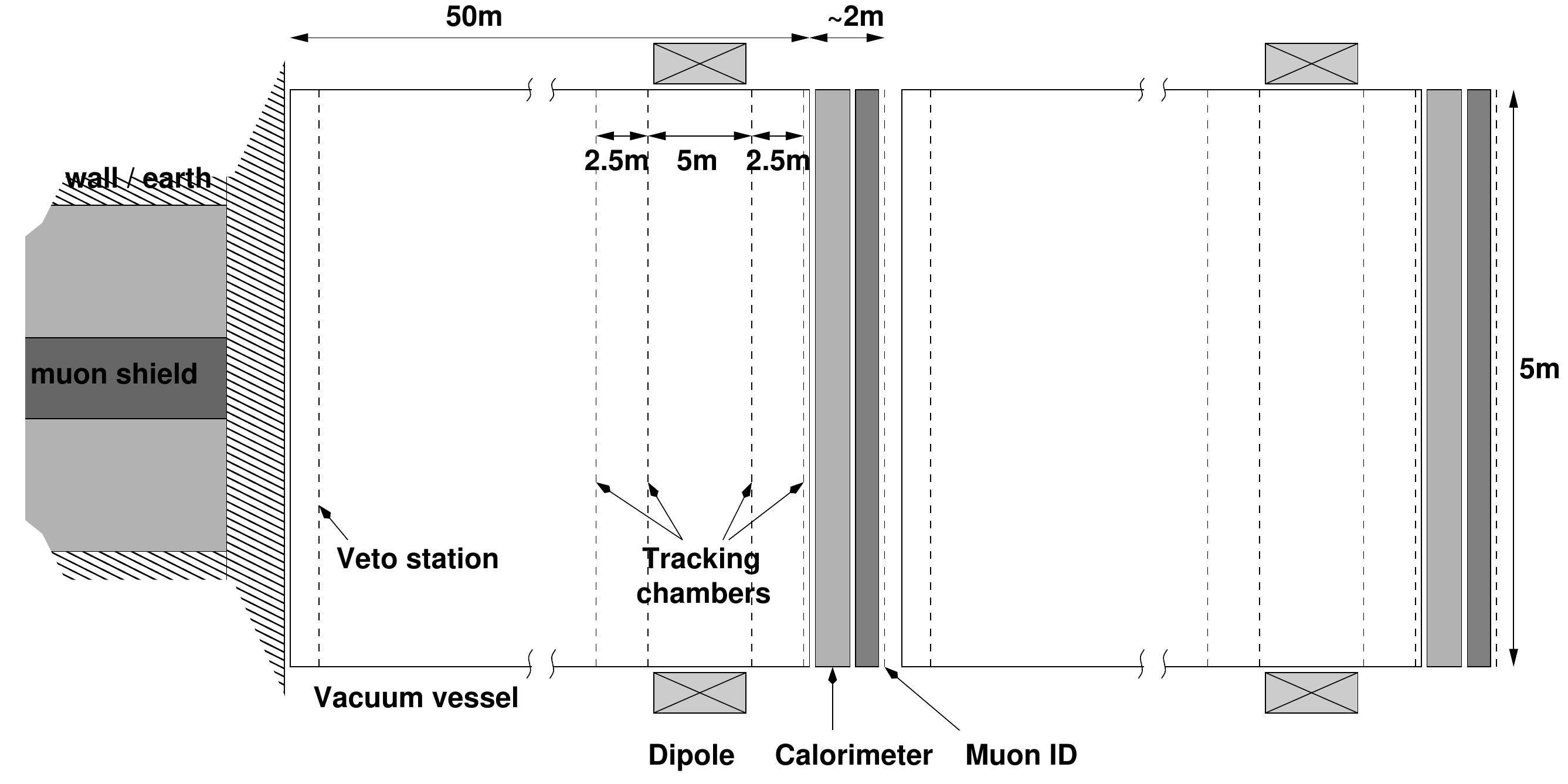}
}
\caption{\small Two-dimensional view of the fiducial volume and detector arrangement.}
\label{Detector}
\end{figure}

\begin{figure}[!tb]
\centerline{
\includegraphics[width=1\textwidth]{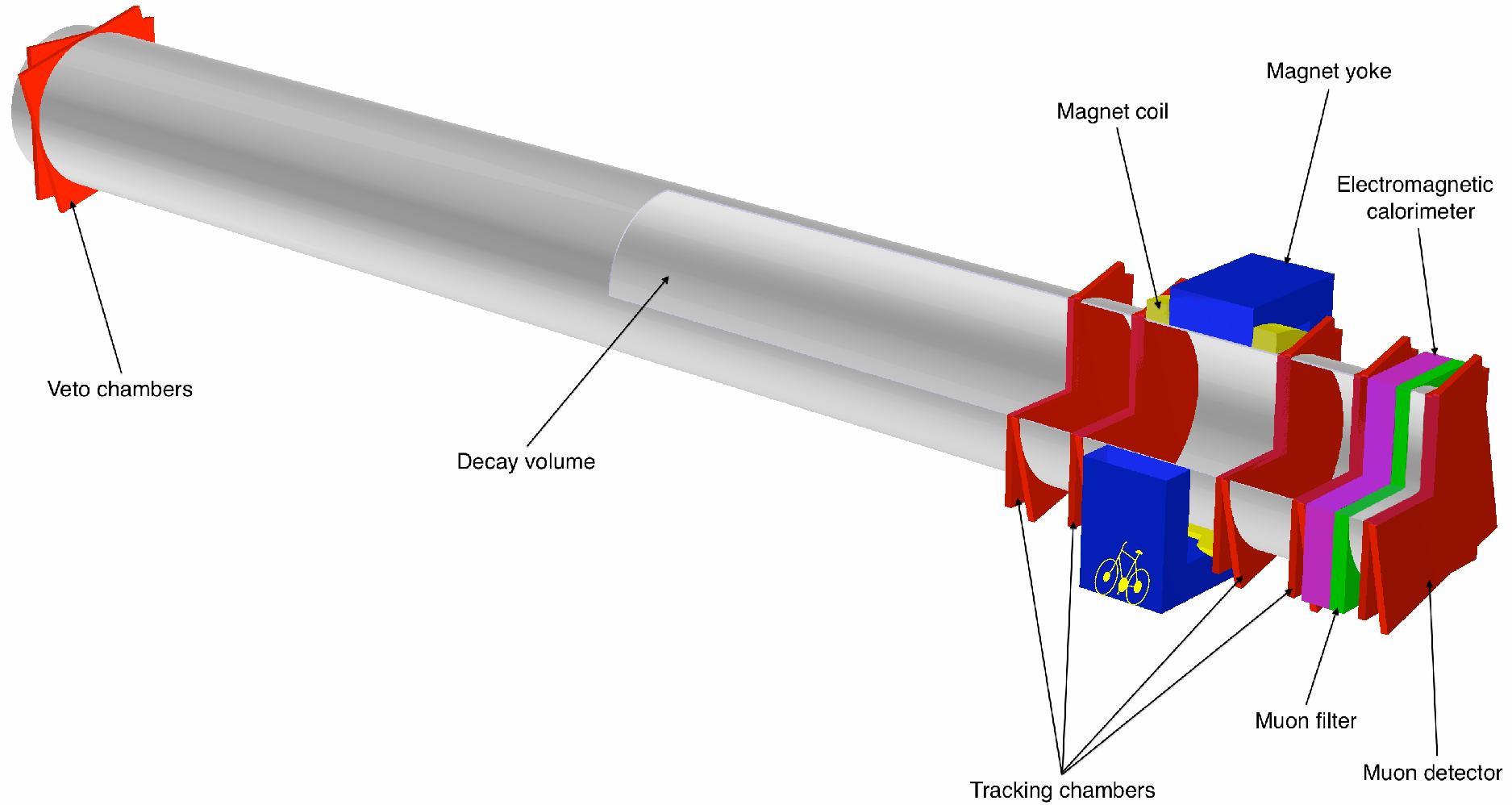}
}
\caption{\small Three-dimensional sketch of the fiducial volume and detector arrangement.}
\label{Detector3D}
\end{figure}

Figure~\ref{Length} shows a scan of the length of the detector for both a single 
detector element and for two longitudinally arranged detector elements. For a given 
HNL lifetime and detector aperture, the number of HNLs decaying in the apparatus 
with the decay products going through the spectrometer saturates 
as a function of the length of the detector. The use of two magnetic spectrometers 
increases the geometric acceptance by 70\% compared to a single element. 
Therefore, the proposed detector will have two almost identical detector elements as 
depicted in Fig.~\ref{Detector}. 
A diagram of a single detector element is also shown in Fig.~\ref{Detector3D}.
 
%%%%%%%%%%%%%%%%%%%%%%%%%%%%%%%%%%%%%%%%%%%%%%%%%%
%%%%%%%%%%%%%%%%%%%%%%%%%%%%%%%%%%%%%%%%%%%%%%%%%%

To reduce to a negligible level the background caused by interactions of neutrinos with the 
remaining air inside the decay volume,  a pressure of less than $\sim10^{-2}$\,mbar 
will be required (see Section~\ref{sec:bg}). Each detector element
therefore consists of a $\sim$50\,m long cylindrical vacuum vessel of 5\,m diameter. 
The first $\sim 40\,$m constitute the decay volume and the subsequent 10\,m are used for the 
magnetic spectrometer. 
The combined calorimeter and muon detector have a length of 2\,m.

%%%%%%%%%%%%%%%%%%%%%%%%%%%%%%%%%%%%%%%%%%%%%%%%%%
%%%%%%%%%%%%%%%%%%%%%%%%%%%%%%%%%%%%%%%%%%%%%%%%%%
%%%%%%%%%%%%%%%%%%%%%%%%%%%%%%%%%%%%%%%%%%%%%%%%%%
%%%%%%%%%%%%%%%%%%%%%%%%%%%%%%%%%%%%%%%%%%%%%%%%%%

\begin{figure}[!tb]
  \vspace*{5mm}
  \centerline{
    \includegraphics[width=0.55\textwidth]{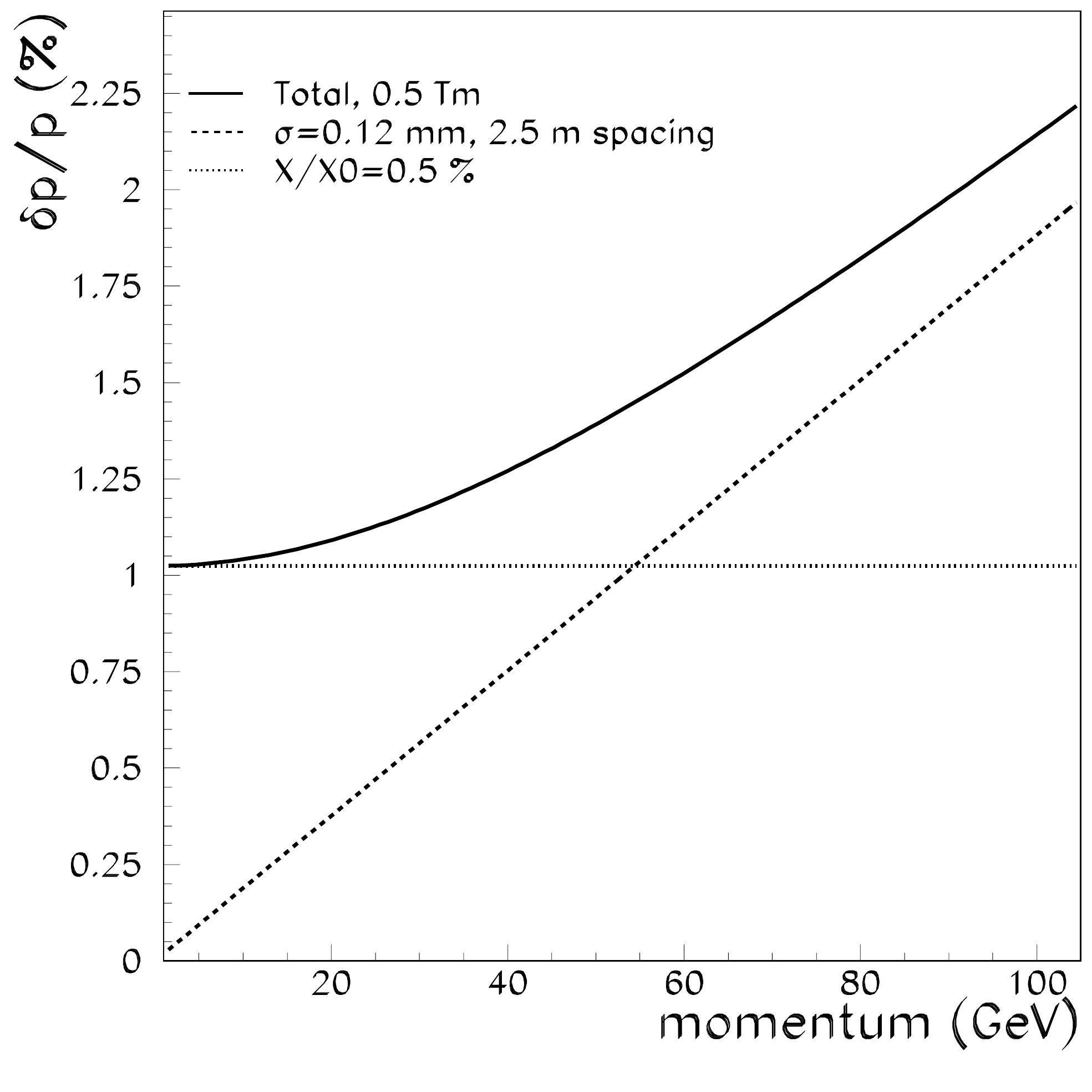}
  }
  \caption{\small Estimated momentum resolution of the spectrometer (solid line) with separate contributions from multiple scattering (dotted line) and chamber resolution (dashed line). }
  \label{magfield}
\end{figure}

The magnetic spectrometer includes a 4\,m long dipole magnet, two tracking 
layers upstream of the magnet, and two tracking layers downstream of the magnet 
(see Fig.~\ref{Detector}). For the required level of vacuum, the tracking chamber 
thickness and resolution are matched to give a similar contribution to the overall 
spectrometer resolution (see Fig.~\ref{magfield}). Using straw tubes with 
$\sim$120\,$\mu$m resolution 
and with 0.5\% $X/X_0$, like those presently being produced 
for the NA62 experiment~\cite{na62}, simulation studies indicate that 2.5\,m is required 
between tracking chambers, giving $\sim 10$\,m length for each magnetic 
spectrometer. 

An electromagnetic calorimeter is located behind each 
vacuum vessel for $\pi^0$ reconstruction and lepton identification. 
The calorimeter material is also part of the muon filter for the muon detector, 
which consists of an iron wall followed by a tracking station. 
An additional tracking station at the beginning of each decay vessel will be 
used to veto charged particles entering the fiducial volume.
These stations will also reject upstream neutrino interactions. 
%The veto station of the downstream detector also performs muon
%identification for the upstream detector. 

%%%%%%%%%%%%%%%%%%%%%%%%%%%%%%%%%%%%%%%%%%%%%%%%%%
%%%%%%%%%%%%%%%%%%%%%%%%%%%%%%%%%%%%%%%%%%%%%%%%%%

\begin{figure}[!tb]
\centerline{
\includegraphics[width=0.95\textwidth]{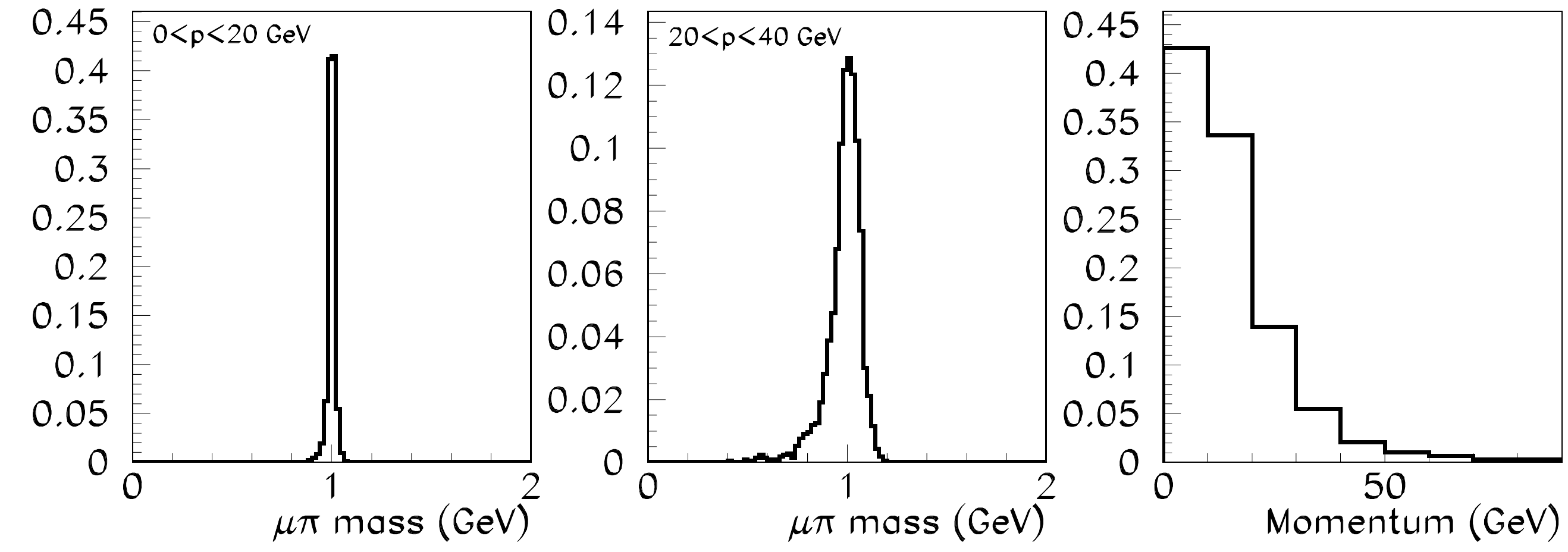}
}
 \caption{\small  Simulated mass resolution for $N\rightarrow\mu^{-}\pi^{+}$ decay products with
 various momenta in the left and middle panels and the momentum spectrum of the decay
 products in the rightmost panel.  The momentum window indicated is for the measured momenta and is required for the harder of
 the two decay products. A HNL with mass 1\,GeV is assumed.}
\label{KL}
\end{figure}

For a mass $M_N = 1$\,GeV, 75\% of the $\mu^{-}\pi^{+}$ decay products 
have both tracks with momentum $p<20\,$GeV.
The momentum and hence mass resolution scales with the integrated field of the
magnets. A 0.5\,Tm field integral results in a mass resolution of 
$\sim40\,$MeV for $p<20\,$GeV tracks~(see Fig.~\ref{KL}).
For a 1\,GeV HNL this provides ample separation between the signal peak and the high 
mass tail of partially reconstructed $K^0_\mathrm{L}\rightarrow \pi^{+}\mu^{-}\nu$ decays. 
Further optimisation of the magnetic field will need to take into account the shape
of the high mass tail from such decays which may enter the signal mass window. 
%%%%%%%%%%%%%%%%%%%%%%%%%%%%%%%%%%%%%%%%%%%%%%%%%%
%%%%%%%%%%%%%%%%%%%%%%%%%%%%%%%%%%%%%%%%%%%%%%%%%%

\subsubsection{Magnet}

A feasibility study of a dipole magnet, similar to the LHCb magnet~\cite{Magnet-TDR}, with a free aperture of almost
16\,m$^2$ and a field integral of $\sim 0.5$\,Tm, has been conducted.

\begin{figure}[!tb]
\centerline{
\includegraphics[width=0.55\textwidth]{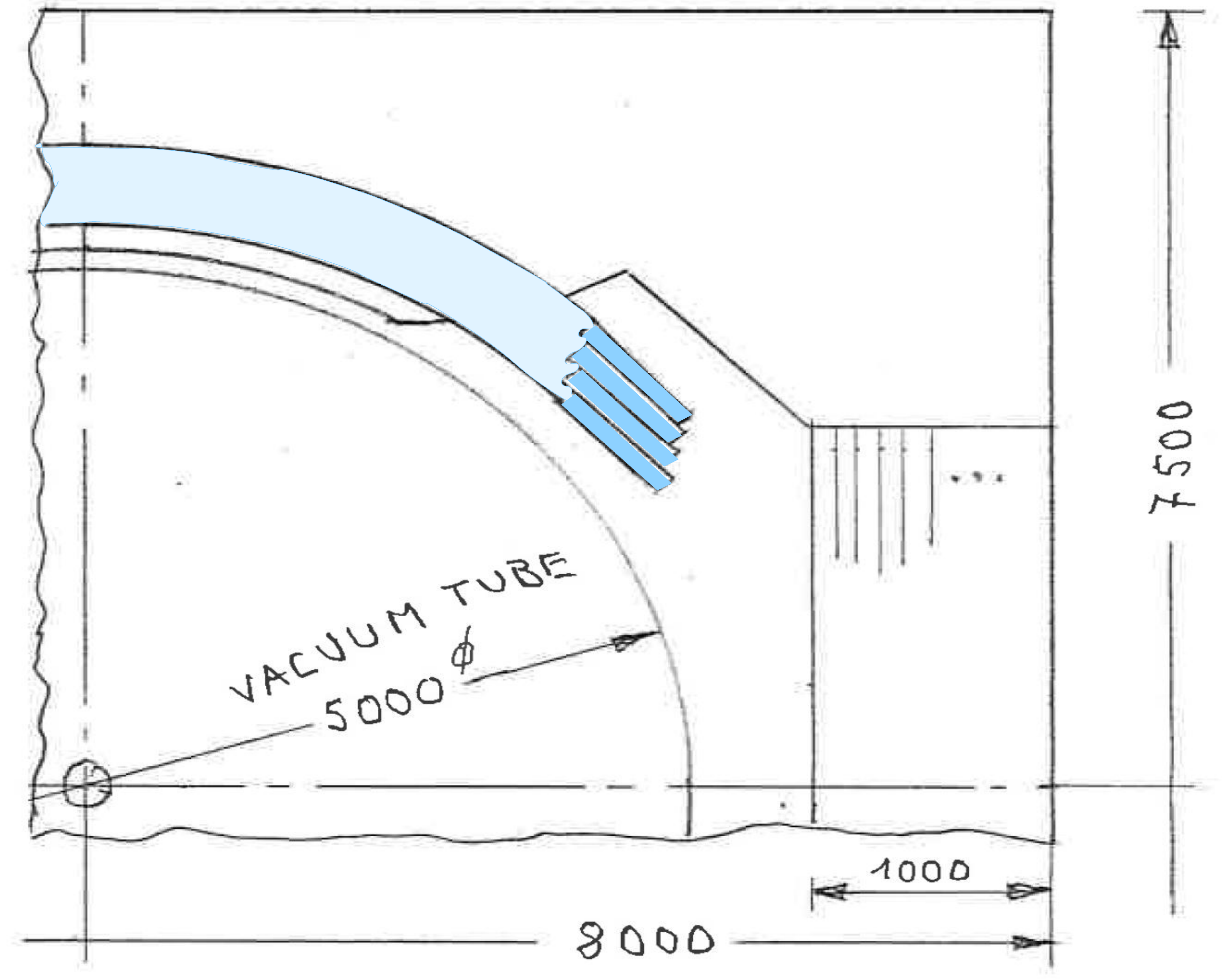}
}
\caption{\small Sketch of a quarter of the dipole magnet viewed along the direction
 of the beam. The Al coil is shown in blue. The units are mm.}
\label{magnet-scketch}
\end{figure}

Figure~\ref{magnet-scketch} shows a sketch
of a magnet which fulfills the requirements of the proposed experiment~\cite{Wilfried}.
%%%%%%%%%%%%%%%%%%%%%%%%%%%%%%%%%%%%%%%%%%%%%%%%%%
%%%%%%%%%%%%%%%%%%%%%%%%%%%%%%%%%%%%%%%%%%%%%%%%%%
With a yoke with outer dimension of $8.0\times 7.5\times 2.5$\,m$^3$, and
two Al-99.7 coils, the proposed magnet provides a
peak field of $\sim 0.2$\,T, and a $\int B dL\approx 0.5$\,Tm over a length of $\sim 5$\,m.
For comparison, the LHCb magnet mentioned above contains $\sim 40~\%$
more iron for its yoke, and dissipates three times more power.

\subsubsection{Synergy with other detectors}
\label{ssec:synergy}

The proposed experiment will require two  $\sim$\,50\,m long, 5\,m diameter
vacuum tanks each of which will be similar to that used in the NA62 experiment~\cite{na62}. 
The $\sim10^{-2}$\,mbar vacuum required to suppress neutrino interactions 
(see Section~\ref{sec:bg}), is several orders of magnitude less demanding 
than the pressure used in the NA62 vacuum tank ($<10^{-5}\,$mbar) and should therefore not represent a technological challenge.

The tracking stations of the magnetic spectrometer must provide good 
spatial resolution and minimise the contribution from multiple scattering. 
The NA62 straw tracker tubes~\cite{na62}, which are manufactured from thin
polyethylene terephthalate (PET), are ideal to meet both of these goals. 
Gas tightness of these tubes has been demonstrated in long term tests and 
the mass production procedure is also well established.

The reconstruction of $\pi^0$ mesons and the identification of electrons would 
be required to reconstruct decays such 
as $N\rightarrow \rho^{+}(\rightarrow\pi^{+}\pi^0) \mu^{-}$ and
$N\rightarrow e^{-}\pi^{+}$. An electromagnetic calorimeter with a modest 
energy resolution will therefore substantially improve the discovery potential 
of the proposed experiment. The LHCb shashlik calorimeter has demonstrated an 
energy resolution of $\frac {\sigma (E)}{E} < 10\%/\sqrt{E} \oplus 1.5\%$, which is 
comparable to the momentum resolution of the proposed magnetic spectrometer in 
the 10 to 20\,GeV energy range.
The shashlik technology also provides an economical solution with fine granularity,
as well as  time resolution better than a few ns, which will be needed
to correlate the calorimeter and tracker information. 

The SPS provides a quasi-continuous flux of protons on target over its extraction period. 
A dead-timeless readout system such as that envisaged for the LHCb Upgrade~\cite{LHCb:upgrade} 
would suffice for the proposed experiment.
This readout system will record data continuously in 25\,ns time-slices. 
The data will be pushed out to read-out boards which in turn will push the data
to a PC-farm to perform the event building.
Tracks with matching times will be combined to form two-prong vertices, 
and events with good vertices will be maintained for further analysis. 
This part of the selection will be executed on-line in the event building 
farm to reduce the data storage rate to a negligible level. 
It is estimated that storage of  $\sim$10\,TB/year would be required.

\section{Background}
\label{sec:bg}

The muon shield described above is designed to stop muons with momenta of up to 400\,GeV, 
thus reducing muon-induced backgrounds to a negligible level.

The rate of charged-current neutrino interactions (CC) occurring at the downstream end of the muon shield is estimated
by extrapolating from a measurement by CHARM~\cite{Charm88}, which used 400\,GeV protons impinging on a Cu target. 
To extrapolate this measurement to the proposed geometry,
the angular and momentum distributions of the neutrinos are simulated using PYTHIA~\cite{Pythia}.
This results in an expected CC rate in the last interaction length of the muon shield of 
$260\times 10^3$ per $2\times 10^{20}$ protons on target if a Cu target were used. 
As a cross-check GEANT~\cite{Geant} is used to simulate the neutrino spectrum 
produced by a 400\,GeV proton beam on a Cu target. This yields a CC rate
four times larger than the estimate based on the CHARM data. 

Replacing the Cu target by a W target lowers the CC rate by $45$\%. 
The neutrinos from the GEANT simulation using a W target are passed to GENIE~\cite{Genie} to simulate 
the CC and neutral-current (NC) neutrino interactions in the muon shield. 
This yields a CC(NC) rate of $\sim 600(200)\times 10^3$ per interaction length per $2\times 10^{20}$ protons on target.
Conservatively, this rate is used to evaluate the background.

Neutrino interactions in the decay volume could be a source of background.
In a decay volume filled with air under atmospheric pressure,
the above rate translates into $\sim 20\times 10^3$ neutrino interactions per 
$2\times 10^{20}$ protons on target.
A pressure in the decay volume of 0.01\,mbar reduces this rate to a negligible level.

\begin{figure}[tb]
   \begin{center}
     \vspace*{-25mm}
   \includegraphics[width=0.48\textwidth]{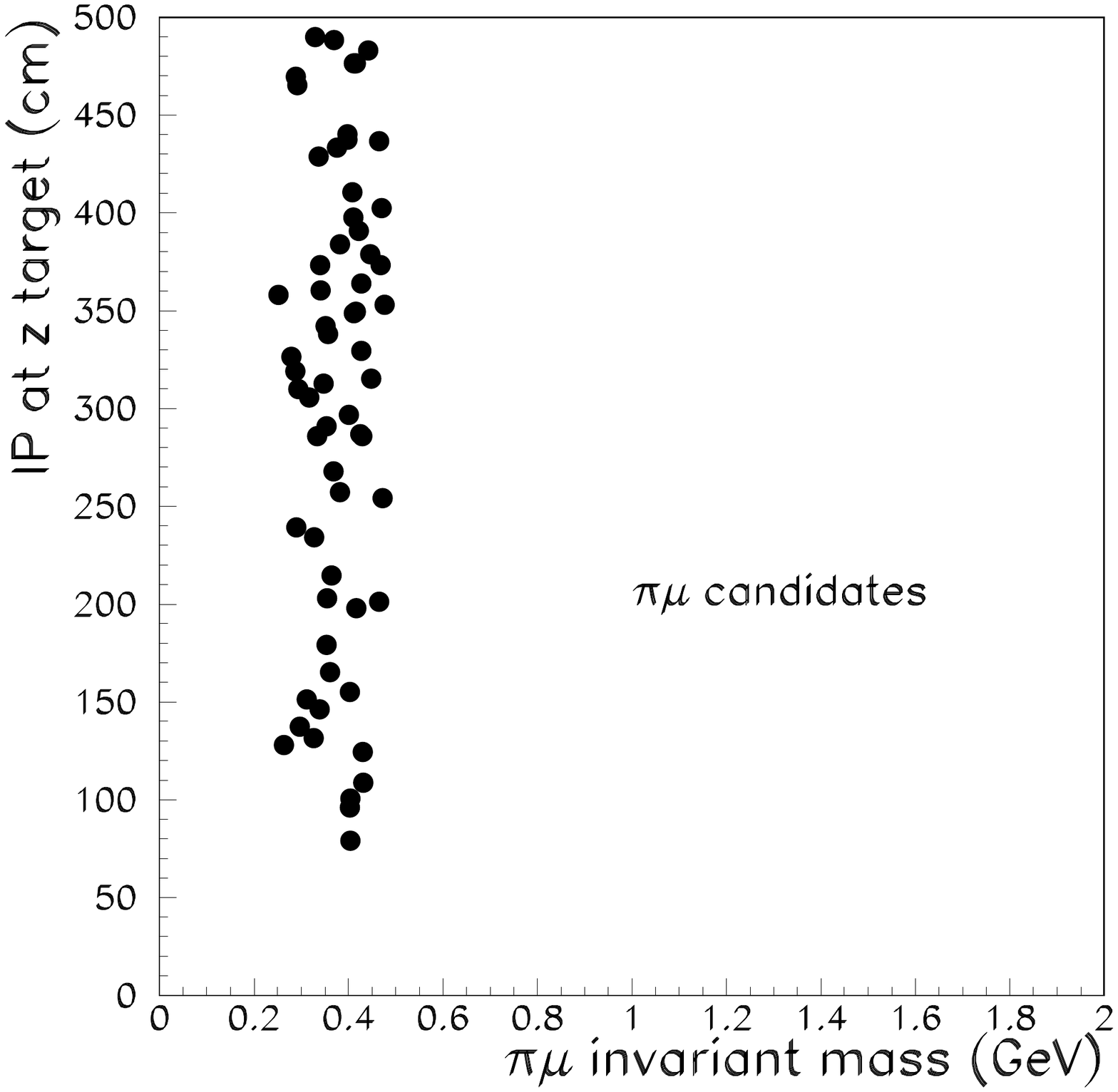}
   \hfill
   \includegraphics[width=0.48\textwidth]{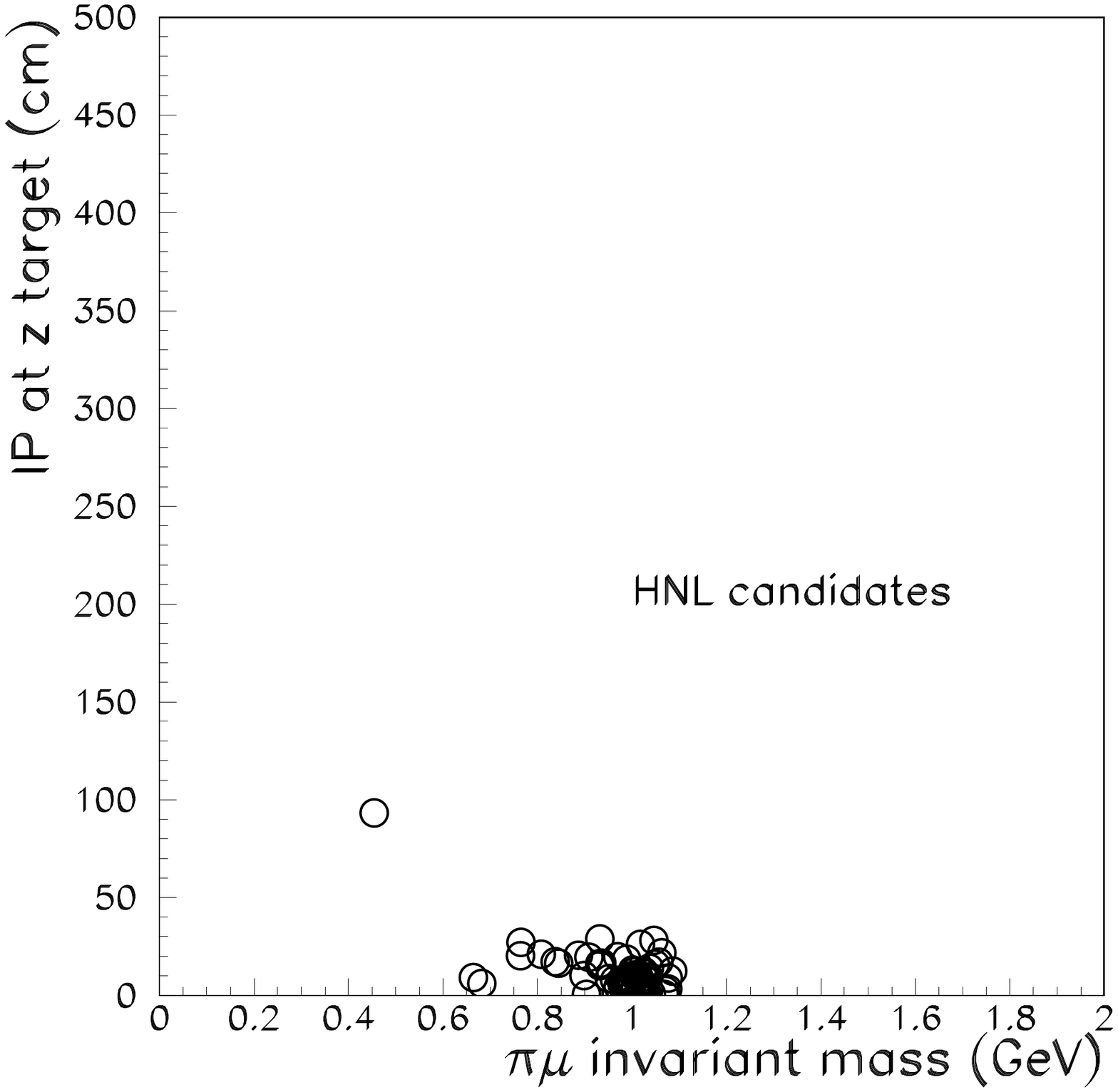}
   \caption{\small Invariant mass of two-prong candidates and their distance of closest approach (IP) to  
     the W target (left) for background events generated with GEANT and GENIE, and (right) for 
     $N\rightarrow\mu^-\pi^+$ with a 1\,GeV invariant mass for the signal.
    }
   \label{fig:KL-mass-IP}
   \end{center}
   \end{figure}

Another source of background is the rate of the neutrino interactions that occur in the muon shield just
upstream of the decay volume.
A combination of GEANT and GENIE is used to predict that in $\sim 10$\% of the neutrino
interactions a $\Lambda$ or $K^0$ will be produced. This prediction is consistent with measurements 
by NOMAD~\cite{strangeness-from-neutrino}. 
In the first 5\,m of the decay volume two-prong vertices are mainly from $\Lambda$ and $K^0_\mathrm{S}$ 
decays. 
For the remaining 35\,m of the decay volume $95$\% of two-prong vertices originate from $K^0_\mathrm{L}$ decays. 
Requiring one of the two decay tracks to be identified as a muon, yields
$\sim 150$ two-prong vertices in $2\times 10^{20}$ protons on target. 
Owing to their different kinematics, the geometrical acceptance of $K^0_\mathrm{L}$ decay products is 
significantly  smaller than that of HNL decay products.
Figure~\ref{fig:KL-mass-IP} shows the
invariant mass of these candidates together with their distance of closest approach (IP) 
when extrapolated back to the W target. 
Requiring IP\,$<1$\,m reduces this background to a handful of candidates for 
$2\times 10^{20}$ protons on target, while the IP of all signal candidates is below 1\,m.
The IP of candidates will also be used to reject possible background induced in 
neutrino interactions in the surrounding material, like the vacuum tank or the floor, and from cosmic rays. 

Backgrounds that originate from neutrino interactions could be vetoed by detecting the
associated activity produced in the associated shower, especially the lepton in CC interactions.
This signature has not been exploited yet in the above yields and could be used to reduce the background to a negligible level.  
More detailed simulation
studies will be undertaken to optimise the position of the veto stations at the entrance of
the decay volume.
If shown to be advantageous, the last interaction lengths of both the muon shield and the muon filter 
could be instrumented in order to detect the products of neutrino interactions. 

\%input{Sensitivity}
\section{Expected event yield}
\label{sec:sens}

The  sensitivity  of the proposed experiment depends on the final states used to reconstruct the 
signal decays and the pattern of couplings between the neutrino generations and the HNLs.
The region in $U_{\mu}^2$ that can be probed by the proposed experiment is conservatively 
estimated by assuming only the decay $N \to \mu^-\pi^+$ is used and considering the production mechanism $D\to\mu^+ N X$.  
The relationship between $U_{\mu}^2$ and $U^2$ is discussed in Ref.~\cite{Ruchayskiy:2011aa}. 
As detailed in Section~\ref{sec:exp_status}, the present strongest limits on $U_{\mu}^2$ 
come from the CHARM experiment~\cite{Bergsma:1985is} and are at the $10^{-7}$ level for a mass 
of 1.5\,GeV and a few times $10^{-6}$ for a mass of 1\,GeV.
The theoretically allowed region for $U^2_{\mu}$ is between $10^{-7}$ and $10^{-11}$. 

The expected number of signal events for the proposed experiment is given by
\begin{equation}
   n_\mathrm{signal} = n_\mathrm{pot} \times 2\chi_{c\overline{c}} \times {\cal{B}}(U_{\mu}^2) \times \epsilon_\mathrm{det}(U_{\mu}^2), 
\end{equation}
where $n_\mathrm{pot}=2\times 10^{20}$ is the total number of protons on target, 
$\chi_{c\overline{c}}=0.45\times 10^{-3}$ is the ratio of the $c\overline{c}$ production 
cross-section with respect to the total cross-section (the factor 2 accounts for the fact that charm is produced in pairs), 
$\epsilon_\mathrm{det}$ is 
the total efficiency and  ${\cal{B}}$ is the product of the production and decay branching fractions of the HNLs. 
The quantity $\epsilon_\mathrm{det}$ can be written as the product of the probability that the HNLs decay in the 
fiducial volume and the trigger and reconstruction efficiencies. 
The probability of decay is estimated with  simulated events for different values of $U_{\mu}^2$ and the efficiencies 
are assumed to be 100\%. 
The dependence on the HNL lifetime introduces a $U_{\mu}^2$ factor in $\epsilon_\mathrm{det}$.
Since the quantities ${\cal{B}}$ and $\epsilon_\mathrm{det}$ each depend on a factor $U_{\mu}^2$, 
this results in an overall $U_{\mu}^4$ dependence. 

Assuming $U^2_{\mu}=10^{-7}$, a HNL mass of 1\,GeV, ${\cal{B}}=8\times 10^{-10}$, $\tau = 1.8\times 10^{-5}$\,s
and $\epsilon_\mathrm{det}=8\times 10^{-5}$,
$\sim12000$ fully reconstructed $N\rightarrow\mu^-\pi^+$ events would be observed, four orders of magnitudes larger than the number that would be
expected in the CHARM experiment. 
Considering a point in the cosmologically favoured  region with $U_{\mu}^2 = 10^{-8}$, $\tau = 1.8\times 10^{-4}$\,s
and $M=1$\,GeV, 
120 fully reconstructed $N\to \mu^-\pi^+$ events would be expected in the proposed experiment. 

The electromagnetic calorimeter allows the reconstruction of decay modes with a neutral pion in the final state 
such as  $N\to \mu^- \rho^+$, where the $\rho^+ \to \pi^+\pi^0$, allowing the signal yields to be doubled. 
Channels with electrons, such as $N\to e^- \pi^+$, could also be studied, allowing a further increase 
in the yields and the parameter $U_e^2$ to be probed.
Other decay channels with a SM neutrino in the final state, e.g.\ $N\to l^-_{1} l^+_{2} \nu$, are more challenging 
to select, since the invariant mass will have a broad distribution but could be separated from the background 
by using particle identification information. 

Both the background levels and the total neutrino flux will be measured by the experiment itself.  
The neutrino flux from charm-meson decays is more difficult to obtain experimentally.  
A study of the $p_\mathrm{T}$ distribution of neutrino events in the calorimeter together with existing 
400\,GeV pCu data~\cite{Charm88} and a detailed simulation will provide a normalisation.
 
In summary, for a HNL mass below 2\,GeV the proposed experiment has discovery potential for the cosmologically 
favoured region with $U^2_{\mu}$ between $10^{-7}$ and a few times $10^{-9}$.

\subsection{Comparison with other facilities}
\label{ssec:sens_other}

Fixed target experiments of the type proposed could be performed using both the Fermilab and KEK proton beams. 
The beams considered are the 800\,GeV and 120\,GeV FNAL beams with $1\times 10^{19}$ protons on 
target and $4\times 10^{19}$ protons on target, respectively; and the KEK 30\,GeV beam with $1\times 10^{21}$ protons on target. 

At FNAL, the 800\,GeV beam would give a similar HNL flux to that of the proposed SPS experiment, 
i.e.\ the lower proton intensity would be approximately compensated by the increase in the charm cross-section at higher energy~\cite{adams2009}. 
However, a significantly longer muon filter would be required due to the higher beam energy, which would be much more challenging, 
leading to a significant loss of acceptance.

The FNAL 120\,GeV beam would have a factor ten lower event yield than in the proposed SPS experiment, while the KEK beam would have 
a factor 1.5\,--\,2 lower yield, the latter estimate has a large uncertainty due to the poor knowledge of  
the charm cross-section at low energy. 

The sensitivity of a colliding beam experiment at the CERN LHC is estimated assuming a luminosity of 1000\,fb$^{-1}$ 
and an energy of 14\,TeV, as is foreseen in three to four years of running for the high luminosity upgrade. 
The HNL decay volume is taken to be located 60\,m away from the interaction region and 50\,mrad off-axis, in order to avoid the LHC beam line.  
The overall HNL event yield would be a factor approximately 200 smaller than in the proposed SPS experiment.

Although masses of the HNLs are expected to be around the GeV-scale, it is possible that they are heavier than $D$ mesons. 
If the HNLs are lighter than $\sim$5\,GeV, they can be produced in beauty hadron decays.  
The most copious HNL production mechanism with $B$ mesons would be the semileptonic $B \to D l N$ decays and the  
total available mass would therefore be restricted to less than $\sim3$\,GeV. 
The reduced cross-section for the production of beauty mesons with respect to charm mesons 
means that the limits that could be derived from a dedicated experiment at the LHC would be about four orders of magnitude 
weaker than those from charm decays. 
Such limits would then  be comparable to the upper limits from the theoretical consideration of the baryon asymmetry.

The SPS at CERN is therefore the ideal facility to conduct the proposed experiment to search for HNLs. 

\section{Conclusion} 
\label{sec:conc}

The proposed experiment will search for New Physics in the largely unexplored domain of new, very weakly 
interacting particles with masses below the Fermi scale. 
This domain is inaccessible to the LHC experiments and to comparable experiments at other existing facilities.

The proposed detector is based on existing technologies and therefore requires no substantial R\&D phase. 
% A suitable collaboration could construct the proposed detector in a few years.
A moderately sized collaboration could construct the proposed detector in a few years.
%The detector could be constructed in a few years. 
%The main challenge of the experiment is 
The design of the beam line is challenging, in particular, the beam extraction 
and beam target, as well as the radiological aspects require further study.  
The solutions proposed are being actively discussed with machine experts.

The impact that a discovery of a HNL would have on particle physics is difficult to overestimate. 
In short, it could solve two of the most important shortcomings of the SM: the origin of the baryon asymmetry of the Universe, 
the origin of neutrino mass.
In addition, the results of this experiment, together with cosmological and astrophysical data, 
could be crucial to determine the nature of dark matter. 

\section*{Acknowledgements}

%We are grateful to the following people for useful discussions and valuable input on the beamline and the target: "
%I would avoid to be explicit about what they did up to now. 
%In fact, only Heinz actually did some simulations but would like to stay low key about it, all the others have been idea "ball planks".
%
%There is some mixup among the people here as well, only the following people should be associated with beamline and target:
%G.~Arduini, M.~Calviani, D.~Grenier, E.~Gschwendtner, H.~Vincke
%
%S.~Gninenko and A. Rozanov are colleagues of Misha, and should have a different text.
%
%We are grateful to the following people for useful discussions:  
%G.~Arduini, M.~Calviani, W.~Flegel, S.~Gninenko, D.~Grenier, E.~Gschwendtner, A.~Rozanov, H.~Vincke.
%We are grateful to 
%G.~Arduini, M.~Calviani, S.~Gninenko, D.~Grenier, E.~Gschwendtner, A.~Rozanov, and H.~Vincke
%for useful discussions.

We are grateful to 
G.~Arduini, M.~Calviani, D.~Grenier, E.~Gschwendtner and H.~Vincke
for useful discussions and valuable input on the beam line and the target.
We would like to thank F.~Rademakers for providing the three dimensional sketch of the experiment.
W.~Flegel is warmly acknowledged for adapting the design of the LHCb magnet to our needs.
We are grateful to S.~Gninenko and A.~Rozanov for stimulating discussions
and to E.~van~Herwijnen for setting up our web site.%\cite{http://snoopy.web.cern.ch/snoopy/}.

%%%%%%%%%%%%%%%%%%%%%%%%%%%%%%%%%%%%%%%%%%%%%%%%%%%%%%%%%%%%%%%%%%%
%\input{refs.tex}

\end{document}